\documentclass[10pt,twocolumn,letterpaper]{article}

\usepackage[pagenumbers]{wacv} 

\usepackage[dvipsnames]{xcolor}
\usepackage{float} 
\usepackage{pifont}
\usepackage{threeparttable}
\usepackage{tablefootnote}

\newcommand{\cmark}{\ding{51}} 
\newcommand{\xmark}{\ding{55}} 
\newcommand{\firsttxt}[1]{\colorbox{red!15}{#1}}
\newcommand{\secondtxt}[1]{\colorbox{orange!15}{#1}}
\newcommand{\thirdtxt}[1]{\colorbox{yellow!15}{#1}}
\newcommand{\first}[1]{\cellcolor{red!15}#1}
\newcommand{\second}[1]{\cellcolor{orange!15}#1}
\newcommand{\third}[1]{\cellcolor{yellow!15}#1}

\definecolor{wacvblue}{rgb}{0.21,0.49,0.74}
\usepackage[pagebackref,breaklinks,colorlinks,allcolors=wacvblue]{hyperref}
\usepackage{xcolor}  
\usepackage{colortbl}       
\usepackage{amsmath}
\usepackage{pifont}
\usepackage{multirow}
\usepackage{tablefootnote}
\usepackage[accsupp]{axessibility}  

\def\wacvPaperID{1908} 
\def\confName{WACV}
\def\confYear{2026}

\title{
Spec-Gloss Surfels and Normal-Diffuse Priors for Relightable Glossy Objects
}

\author{
{Georgios Kouros \qquad Minye Wu \qquad Tinne Tuytelaars} \\
{Department of Electrical Engineering (ESAT), KU Leuven, Belgium} \\
{\tt\small \{georgios.kouros,minye.wu,tinne.tuytelaars\}@esat.kuleuven.be}
}

\begin{document}
\maketitle
\begin{abstract}

Accurate reconstruction and relighting of glossy objects remains a longstanding challenge, as object shape, material properties, and illumination are inherently difficult to disentangle. Existing neural rendering approaches often rely on simplified BRDF models or parameterizations that couple diffuse and specular components, which restrict faithful material recovery and limit relighting fidelity. We propose a relightable framework that integrates a microfacet BRDF with the specular-glossiness parameterization into 2D Gaussian Splatting with deferred shading. This formulation enables more physically consistent material decomposition, while diffusion-based priors for surface normals and diffuse color guide early-stage optimization and mitigate ambiguity. A coarse-to-fine environment map optimization accelerates convergence, and negative-only environment map clipping preserves high-dynamic-range specular reflections. Extensive experiments on complex, glossy scenes demonstrate that our method achieves high-quality geometry and material reconstruction, delivering substantially more realistic and consistent relighting under novel illumination compared to existing Gaussian splatting methods. The source code is available at \url{https://github.com/gkouros/SpecGloss-GS}.

\end{abstract}

\section{Introduction} \label{sec:intro}

\begin{figure}
    \centering
    \includegraphics[width=\linewidth, trim=0 0cm 0 0,clip]{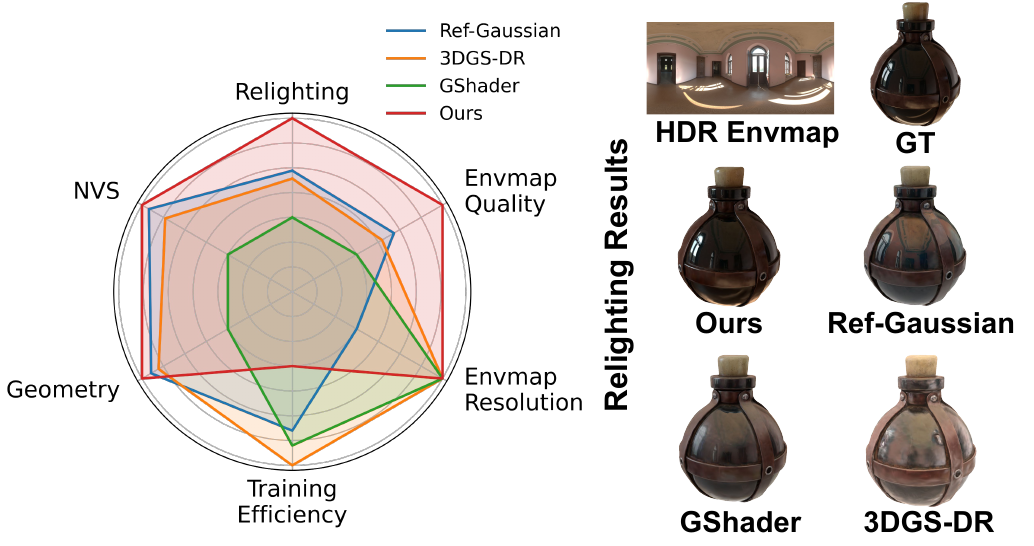}
    \caption{
    Our method achieves high-quality reconstruction with realistic materials and relighting through priors and a disentangled rendering model.}
    \label{fig:introduction:teaser}
    \vspace{-3mm}
\end{figure}

Incorporating real-world objects into virtual environments is a fundamental task in applications such as digital twins and virtual reality. A major challenge arises in accurately reconstructing objects from the physical world and reproducing realistic lighting effects within specific virtual settings. This difficulty is particularly pronounced for highly reflective objects, where the lack of multi-view consistency renders traditional approaches, such as multi-view stereo (MVS), ineffective and makes reliable reconstruction a non-trivial problem.

To overcome these limitations, researchers have extensively explored the field of inverse rendering. Recent advances integrate diverse scene representations with various differentiable rendering models, enabling the reconstruction of highly reflective objects from multi-view images. 

The choice of scene representation and rendering model introduces unique trade-offs in performance. For instance, approaches based on Radiance Fields \cite{mildenhall2020nerf, SunSC22dvgo, barron2022mip360} or 3D Gaussian Splatting (3DGS) \cite{kerbl20233dgs} often fail to accurately capture object surfaces, thereby limiting their ability to model surface reflections. In contrast, methods employing Signed Distance Functions (SDFs) \cite{wang2021neus, ge2023ref, liu2023nero, liang2023envidr} provide precise surface geometry, but their high computational cost poses challenges for real-time rendering.  Another critical component of inverse rendering is the rendering model, which decomposes object materials into multiple attributes and employs predefined functions to synthesize appearance from target viewpoints. To achieve differentiability, various methods approximate the rendering equation integrals differently. For example, Ref-NeRF \cite{verbin2022refnerf} uses an Integrated Directional Encoding, GaussianShader~\cite{jiang2024gshader} employs specular GGX \cite{Walter2007Microfacet}, Ref-GS \cite{zhang2024refgs} adopts a spherical-mip encoding, and 3DGS-DR \cite{ye2024gsdr} leverages deferred rendering. While these diverse formulations yield varying reconstruction quality, they share a fundamental limitation: the inverse solving process suffers from an inherent ambiguity, where multiple material-lighting combinations can produce identical appearances. Consequently, reconstructed material attributes and recovered lighting often deviate from ground truth values, leading to degraded rendering quality under novel lighting conditions. This is not a new observation, but a long-standing open problem going back to early work on  intrinsic images~\cite{BarrowTenenbaum1978}, highlighted more recently in \cite{Kouros_2024_CVPR}.

In this work, we 
reduce ambiguity and improve relighting quality. We start by investigating the relationship between rendering models and relighting effects. Specifically, we analyze 
the coupling between object materials and various BRDF parameterization methods. Based on this, we select a specular-glossiness (Spec-Gloss) \cite{Lagarde2014MovingPBR} material parameterization. 
We find that this decomposition scheme 
extracts reasonable material attributes more effectively compared to other rendering models. Instead of 3DGS, we adapt this rendering model for a 2DGS-based scene representation~\cite{huang20242dgs}, enabling high-quality reconstruction by providing high accuracy surface modeling. We leverage priors from diffusion models to steer the optimization of geometry and materials and to reduce ambiguity in the inverse solving process, resulting in more realistic material attributes and environment lighting. We propose an efficient environment map training strategy that, through a coarse-to-fine approach, accelerates scene reconstruction, enables high-resolution environment map training, and enhances reconstruction quality while assisting the optimization of object material properties. Experimental results demonstrate that our method achieves high-quality geometry and material reconstruction, while delivering more photorealistic relighting results, significantly surpassing those of previous approaches. In summary, our contributions include:
\begin{itemize}
    \item 
    We propose the integration of a microfacet BRDF with Spec-Gloss parameterization into the 2D Gaussian Splatting framework. This enables more accurate material decomposition by decoupling material attributes.
    
    \item
    We incorporate priors from diffusion-based predictors for surface normals and diffuse color to reduce geometry–material–lighting ambiguity and to guide early-stage optimization.
    
    \item 
    We introduce a coarse-to-fine training strategy for the environment map, improving both training efficiency and specular reconstruction by gradually increasing lighting resolution.
\end{itemize}
\section{Related Work} \label{sec:related}

\begin{figure*}[ht]
    \centering
    \includegraphics[width=\linewidth]{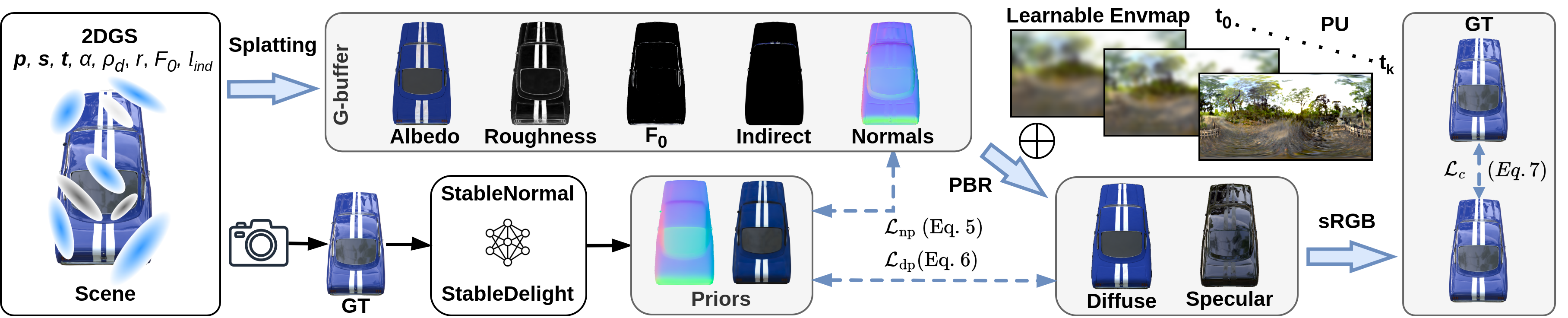}
    \caption{Overview of our framework built on 2DGS~\cite{huang20242dgs}. Gaussian splats rasterize to a G buffer of albedo, roughness, $F_0$, indirect color, and surface normals. A differentiable prefiltered environment cubemap with mipmaps provides lighting in a physically based deferred renderer. The HDR environment map is learned in a coarse-to-fine manner. Supervision uses an sRGB photometric loss between shaded output and ground truth (GT), plus normal and diffuse priors that reduce ambiguity between geometry, materials, and lighting.}
    \label{fig:architecture}
    \vspace{-3mm}
\end{figure*}

\subsection{Neural Rendering for Glossy Scenes}
Reconstructing glossy scenes is hard because specular cues are high frequency and view dependent, which creates multi-view inconsistencies and entangles geometry, materials, and light. NeRF \cite{mildenhall2021nerf}, DVGO \cite{SunSC22dvgo}, and NeuS \cite{wang2021neus} advance geometry and novel view synthesis but struggle with specular fidelity or slow training. Ref-NeRF \cite{verbin2022refnerf} adds a simplified BRDF and implicitly learns reflected radiance, but remains slow and prevents explicit relighting. 3DGS \cite{kerbl20233dgs} improves efficiency and quality but is not relightable and has no explicit surface normals in the representation. GaussianShader \cite{jiang2024gshader} extends 3DGS with a simple specular term, yet yields noisy normals, which hurts geometry, materials, and lighting recovery. R3DG \cite{R3DG2024} improves geometry by attaching BRDF to points and using ray-based visibility, but per-point queries scale poorly on dense point clouds. 3DGS-DR \cite{ye2024gsdr} scales better and adopts deferred shading with normal propagation from the shaded buffer to the primitives, which stabilizes geometry, but it does not use a full microfacet BRDF, so disentanglement and relighting fidelity remain limited. 2DGS \cite{huang20242dgs} introduces view-consistent planar disks with explicit normals, better suited to reconstructing glossy objects. Subsequent works \cite{zhang2024refgs,yao2025refgaussian,kouros2025rgsdr,tong2025gs2dgs} adopt 2DGS with pixel-level deferred shading and differentiable environment lighting, alleviating geometry issues and scaling constraints. We build on this line of work and use 2DGS with deferred shading as our foundation.

\subsection{Inverse Rendering}

Inverse rendering seeks materials and illumination that explain multi-view images. Early neural methods incorporated BRDFs and environment lighting into NeRF-style pipelines but were slow, unable to relight, or required prior geometry or lighting \cite{verbin2022refnerf, kouros2023refdvgo, zhang2021nerfactor, srinivsan2021nerv}. SDF-based approaches improved geometry and material quality but were costly \cite{liang2023envidr, liu2023nero, fan2023factoredneus, ge2023ref}. Gaussian splats with forward shading blurred specular cues and hindered disentanglement \cite{jiang2024gshader,R3DG2024}, while deferred pipelines separated rasterization from shading and improved stability \cite{ye2024gsdr}. Recent 2DGS work coupled deferred shading with image-based lighting, preserving sharp highlights and improving optimization \cite{zhang2024refgs, yao2025refgaussian, tong2025gs2dgs, kouros2025rgsdr}.

Material parameterization and BRDF remain central. Disney’s metallic-roughness is intuitive but ties specular reflectance to base color, harming identifiability \cite{burley2012disney}. Specular-glossiness treats specular reflectance as free, alleviating entanglement. Hybrid parameterizations \cite{jiang2024gshader, yao2025refgaussian} introduce specular tint but are incompatible with standard graphics workflows. ReCap \cite{li2025recap} adopts Spec-Gloss in 3DGS but needs multiple illuminations. We also adopt Spec-Gloss in a split-sum approximation with 2DGS and deferred shading for better surface reconstruction, incorporating diffuse color and normal priors. Unlike \cite{ye2024gsdr, yao2025refgaussian}, we avoid bounding positive HDR radiance, using negative-only clipping to preserve bright highlights and improve material-lighting disentanglement. To reduce training time, we propose coarse-to-fine cubemap upsampling paralleling frequency-progressive spherical-harmonics lighting \cite{liu2017progressivesh}.


\subsection{Inverse Rendering Priors}
As already established, inverse rendering is ill-posed since many material–lighting pairs can explain the same images, which harms downstream scene editing tasks. To reduce this ambiguity, prior work adds structure in different ways, such as constraining illumination \cite{srinivsan2021nerv}, using data-driven BRDF priors \cite{zhang2021nerfactor, liang2023envidr}, supplying pretrained or externally supervised geometry \cite{liu2023nero,fan2023factoredneus}, seeding materials with predictors \cite{du2024gsid}, or injecting surface-normal and depth priors during optimization \cite{tong2025gs2dgs}. We follow this line by using learned normal and diffuse color priors through the diffusion models StableNormal \cite{ye2024stablenormal} and StableDelight \cite{stabledelight2025} for early stabilization and to steer the optimization in the right direction.

\section{Methodology} \label{sec:methodology}


Our method integrates the Specular-Glossiness BRDF rendering model with a 2DGS-based scene representation and a pixel-level deferred shading technique~(\S~\ref{sec:methodology:brdf}). To balance speed and quality, we progressively upsample the environment light cubemap in a coarse-to-fine manner~(\S~\ref{sec:methodology:env}). During scene reconstruction, we leverage diffusion priors on normals and material properties to reduce ambiguities and better disentangle materials from illumination~(\S~\ref{ssec:methodology:priors}). The proposed methodology is illustrated in Fig.~\ref{fig:architecture}.  


\subsection{Preliminary}
\noindent \textbf{Surfel-based Scene Representation}. The main limitations of 3DGS \cite{kerbl20233dgs} are its multi-view geometric inconsistencies and the absence of explicit surface normals, both of which hinder inverse rendering of glossy objects. To address this, we build on 2DGS~\cite{huang20242dgs}, which collapses 3D Gaussians into view-consistent planar disks and exposes explicit surface normals aligned with their shortest axis. This design yields thinner surfaces and enables more stable optimization, making it better suited for high-frequency specular effects.

A 2D splat is defined by its center $\mathbf{p}_k$, two principal tangential vectors $\mathbf{t}_u, \mathbf{t}_v$, a scaling vector $S=(s_u, s_v)$. The orientation of the 2D primitive is given by a $3\times3$ rotation matrix $\mathcal{R}=[\mathbf{t}_u,\mathbf{t}_v,\mathbf{t}_w]$ and its explicit normal is defined as $\mathbf{n}=\mathbf{t}_u \times \mathbf{t}_v$. The 2D Gaussian value of a point $\mathbf{u}=(u,v)$ in the $uv$ local tangent space can be determined via $\mathcal{G}(u,v)=exp(-(u^2+v^2) / 2)$. The position $p_k$, scaling $S$, and rotation $R$ are learnable sets of parameters and combined with a learnable opacity value $\alpha$, can be used to rasterize various properties such as color by blending ordered points overlapping the pixel:
\begin{equation} \label{eq:alpha_blending}
c(\mathbf{x}) 
= \sum_{i=1}^{n} 
   \mathbf{c}_{i}\,\alpha_{i}\,\mathcal{\hat{G}}_{i}\bigl(\mathbf{u}(\mathbf{x})\bigr)
   \prod_{j=1}^{i-1}\!\Bigl(1 - \alpha_{j}\,\mathcal{\hat{G}}_{j}\bigl(\mathbf{u}(\mathbf{x})\Bigr),
\end{equation}
where $c$ refers to any Gaussian-level parameter, and $\mathcal{\hat{G}}$ is a low-pass-filtered $\mathcal{G}$. More details can be found in \cite{huang20242dgs}.\\

\subsection{BRDF Parameterization and Rendering} \label{sec:methodology:brdf}
\noindent \textbf{Attribute Disentanglement.}  
A core component of inverse rendering is the bidirectional reflectance distribution function (BRDF), which governs how surface materials interact with incoming light. In practice, this requires choosing both a microfacet model and a parameterization for the material terms. Most recent methods adopt the GGX microfacet model due to its balance of realism and differentiability.

The widely used Disney BRDF~\cite{burley2012disney} employs a metallic–roughness (MR) parameterization with base color $b \in \mathbb{R}^3$, metallic $m \in [0,1]$, and roughness $r \in [0,1]$. The specular reflectance is implicitly defined via:
\begin{equation}
    F_0(m, b) = (1 - m) \cdot 0.04 + m \cdot b,
\end{equation}
which blends a dielectric constant with the base color depending on the metallic value. While intuitive and physically plausible, this formulation entangles diffuse color and specular reflectance, making it ill-suited for inverse rendering. The resulting ambiguity hinders material-lighting disentanglement and degrades relighting and editability.

In contrast, the specular-glossiness (SG) parameterization decouples diffuse and specular components by explicitly learning the specular reflectance $F_0$ as a free RGB parameter. This enables a clean separation of appearance into diffuse albedo $\rho_d$, specular reflectance at normal incidence $F_0$, and roughness $r$, improving the identifiability of material properties, especially under glossy reflections.

Another option is the hybrid metallic-roughness-specular (MRS) parameterization, consisting of base color $b$, metallic $m$, roughness $r$ and specular tint $k_s$, where $k_s$ decouples $b$ from $F_0$. The MRS parameterization introduces additional degrees of freedom, making the inverse problem more underdetermined and prone to ambiguities, while its adoption in practical rendering pipelines remains limited compared to the widely supported Disney or SG models.

\begin{figure}[t]
    \centering
    \includegraphics[width=\linewidth,trim=0.1cm 0cm 0 0,clip]{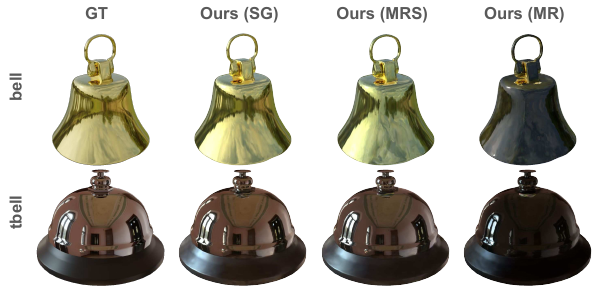}
    \caption{Relighting of bell and tbell scenes from the Glossy Synthetic dataset \cite{liu2023nero} with different material parameterizations integrated into our methodology, demonstrating the improved material-lighting disentanglement of the Spec-Gloss variant.}
    \label{fig:methodology:brdf}
    \vspace{-3mm}
\end{figure}

Based on these considerations, we adopt the SG parameterization in our method. Its simplicity, compatibility with inverse rendering, and ability to produce disentangled material outputs make it a stronger candidate for editable and relightable scene reconstruction, as shown in \cref{fig:methodology:brdf}.

\vspace{3mm}\noindent \textbf{IBL with Split-Sum Approximation.}
To render glossy surfaces, we adopt the split-sum approximation for image-based lighting (IBL)~\cite{Karis2013ibl}, which decomposes illumination into diffuse and specular terms. The diffuse term $L_d(\mathbf{n})$ is obtained by sampling a low-frequency irradiance map,
$L_d(\mathbf{n}) \approx \text{PrefilterEnv}(\mathbf{n}, r^{\max})$, where $\text{PrefilterEnv}(\cdot, r)$ samples a mipmapped environment map prefiltered over roughness $r$ and $r^{\max}$ denotes the maximum roughness (coarsest mip level). For the specular term, we prefilter the environment map with GGX importance sampling~\cite{trowbridge1975average}, yielding a \emph{direct} specular IBL component
$L_{\text{dir}}(\mathbf{v},\mathbf{n},r) \approx \text{PrefilterEnv}(\boldsymbol{\omega}_r, r)$, where $\mathbf{v}$ is the viewing direction, $\boldsymbol{\omega}_r$ is the reflection direction about the surface normal $\mathbf{n}$, and $r$ is the roughness. Indirect specular interreflections are modeled by a low-frequency spherical-harmonics term $L_{\text{ind}}(\boldsymbol{\omega}_r)$ (see supplementary). A visibility term $V \in \{0,1\}$ blends direct and indirect specular illumination:
\begin{equation}
    L_s(\mathbf{v}, \mathbf{n}, r)
    = V \, L_{\text{dir}}(\mathbf{v}, \mathbf{n}, r)
    + (1 - V)\, L_{\text{ind}}(\boldsymbol{\omega}_r).
\end{equation}
Following the split-sum formulation, the remaining GGX BRDF integration is precomputed in a 2D lookup table $\text{BRDF}_\text{LUT}(r,\mathbf{n}\cdot\mathbf{v})$, which returns scale and bias coefficients $\beta_1$ and $\beta_2$ for the Fresnel term. The outgoing radiance is then
\begin{equation}\label{eq:specglossinmethod}
    L_o(\mathbf{v}, \mathbf{n}) =
    \rho_d\, L_d(\mathbf{n})
    + \bigl(F_0\, \beta_1 + \beta_2 \bigr) \odot L_s(\mathbf{v}, \mathbf{n}, r),
\end{equation}
where $\rho_d$ is the diffuse albedo, $F_0$ is the specular reflectance at normal incidence, and $\odot$ denotes element-wise multiplication.


\vspace{2mm}
\noindent \textbf{Deferred Shading.} 
Similar to \cite{ye2024gsdr,tong2025gs2dgs,yao2025refgaussian}, we adopt pixel-level deferred shading and alpha-blend the per-Gaussian attributes (albedo $\rho_{d,i}$, roughness $r_i$, specular reflectance $F_{0,i}$, and surface normals $\mathbf{n}_i$) using \cref{eq:alpha_blending} to construct the screen-space G-buffer with dense per-pixel maps $I_d$, $R$, $F_0$, and $N$. Physically-based BRDF shading is then applied at the pixel level. This decoupling of geometry and material estimation from shading leads to sharper highlights, better supervision signals during training, and robustness to view-dependent specularities compared to forward rendering.

\subsection{Coarse-to-fine HDR Lighting}\label{sec:methodology:env}
We represent scene illumination using a learnable HDR environment map $E$, parameterized as a cube mipmap. During training, the cube map provides efficient prefiltered lookups for the split-sum approximation based on \cite{Laine2020diffrast}. 

\begin{figure}
    \centering
    \includegraphics[width=\linewidth]{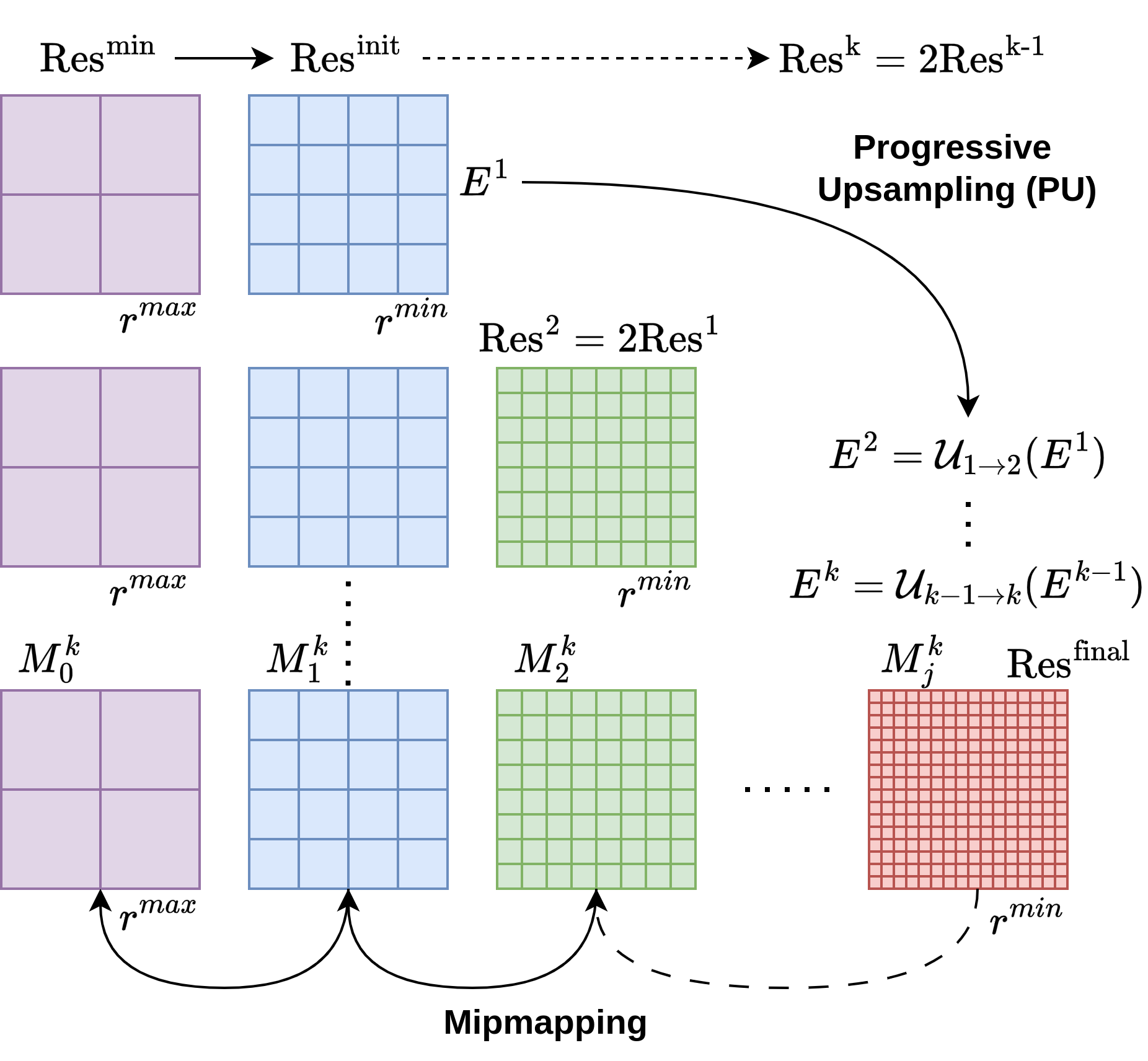}
    \caption{The progressive upsampling (PU) and mipmapping operations of the environment map in 2D for ease of visualization.}
    \label{fig:methodology:pu}
    \vspace{-2mm}
\end{figure}

\noindent\textbf{Progressive Upsampling (PU).}
To mitigate the computational bottleneck of mipmap generation in the split-sum formulation, we adopt a coarse-to-fine training strategy. The cubemap is initialized at a low per-face resolution, $\text{Res}^\text{init}$, and progressively doubled at fixed iteration intervals until $\text{Res}_\text{final}$. At each step, the environment map is bilinearly interpolated, and the cube mipmaps are rebuilt down to $\text{Res}_\text{min}$, reducing early training cost, preventing lighting from absorbing material errors, and ensuring robust coarse-scale convergence before high-frequency lighting is introduced.
At the same time, we avoid redundant smoothing operations from prior works \cite{zhang2021nerfactor,zhang2021physg,Ummenhofer2024OWL}. Formally, given $E^k \in \mathbb{R}^{6\times \text{Res}^k \times \text{Res}^k \times 3}$, an HDR cubemap at stage $k=1,...,K$, we perform progressive upsampling of $E$, defined as $ E^{k+1} = \mathcal{U}_{\text{r}_{k} \rightarrow \text{r}_{k+1}} (E^k)$, 
where $\text{Res}_\text{k+1} = 2 \times \text{Res}^\text{k}$ and $\mathcal{U}$ denotes the bilinear upsampling operation. Then we rebuild the prefiltered cube mip pyramid $\{M_j^k\}_j = \text{PMREM}(E^k)$, where PMREM stands for "Prefiltered, Mipmapped Radiance Environment Map" and is an operation that successively downsamples the cubemap with average pooling until reaching the minimum resolution $\text{Res}^\text{min}$ from the highest resolution $\text{Res}^\text{k}$, thus generating a cube mipmap, as shown in \cref{fig:methodology:pu}.

\begin{figure}[ht]
    \centering
    \includegraphics[width=\linewidth,trim=0.2cm 0 0 0,clip]{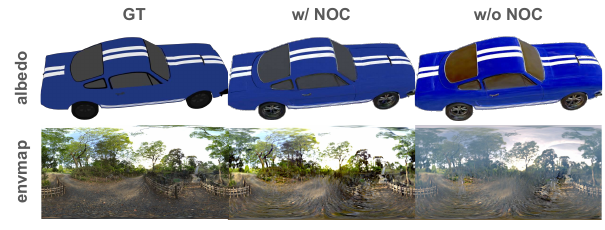}
    \caption{Negative-only clipping (NOC) preserves specular highlights and improves illumination recovery. “w/o NOC” denotes standard clipping to the [0,1] range, where both negative values and values above 1 are clipped.}
    \label{fig:ablation:noclip}
\end{figure}

\noindent\textbf{Negative-only Clipping (NOC).}
In reality, environment lights contribute to dynamic contrast, so its range should not be restricted.
We enforce non-negativity on the HDR envmap and leave positives unbounded:
\(E\leftarrow\max(0,E_{\text{raw}})\) each iteration. Prior work clamps 
\(E_{\text{raw}}\in[0,1]\)~\cite{ye2024gsdr,yao2025refgaussian}, which underexposes lighting and deteriorates material-light disentanglement. NOC preserves HDR peaks required for sharp, bright specular reflections and avoids systematic underexposure, as shown in \cref{fig:ablation:noclip}.

\subsection{Ambiguity Suppression} \label{ssec:methodology:priors}

Inverse rendering from multi-view images is highly under-constrained. To stabilize optimization and improve the disentanglement of materials and illumination, we incorporate priors on geometry and appearance.\\


\noindent\textbf{Surface Normal Prior.}
\begin{figure}[ht]
    \centering
    \includegraphics[width=\linewidth,trim=0 0.05cm 0 0,clip]{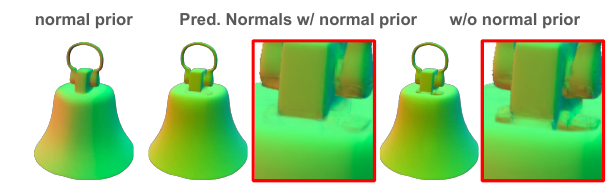}
    \caption{The surface normal prior reduces ambiguity from view-dependent effects such as interreflections, yielding more accurate predicted normals and improved reconstruction quality.}
    \label{fig:ablation:normal_prior}
\end{figure}
We leverage StableNormal \cite{ye2024stablenormal} to provide per-pixel estimates of surface normals. These normals act as soft constraints on the reconstructed G-buffer, encouraging consistent geometry (see \cref{fig:ablation:normal_prior}) and sharper specular highlights. Specifically, the normal prior $\tilde{N}$ is used as pseudo-GT for the predicted normals $\hat{N}$ and the normals derived from the gradient of the surface depth $\hat{N}_D$. The normal prior regularization term is defined as
\begin{equation}
    \mathcal{L_\text{np}} = (1 - \hat{N}^T \tilde{N}) + (1-{\hat{N}_D}^T \tilde{N}).
\end{equation}

\begin{figure}[ht]
    \centering
    \includegraphics[width=\linewidth,trim=0.2cm 0.5cm 0 0,clip]{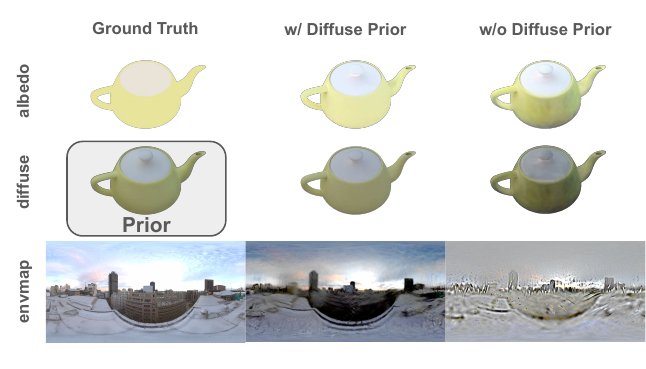}
    \caption{The diffuse color prior reduces ambiguity between albedo and illumination and improves their disentanglement.}
    \label{fig:ablation:diffuse_prior}
\end{figure}

\noindent\textbf{Diffuse Color Prior.}
We incorporate a diffuse color prior based on StableDelight \cite{stabledelight2025}, a diffusion model that removes specular reflections from natural images, thus generating a powerful prior for the diffuse color of a scene, as demonstrated in \cref{fig:ablation:diffuse_prior}. This prior provides a strong initialization, guiding the network toward realistic color decomposition in the early phase of training.  While the prior offers useful guidance, its accuracy may be surpassed by the results obtained after scene optimization. 
To exploit its benefits while mitigating these drawbacks, we apply it only as a soft prior at the beginning of training and then disable it, allowing the model to optimize solely based on the RGB ground truth. The diffuse color prior term is defined as
\begin{equation}
    \mathcal{L}_\text{dp} = \lambda_\text{dp}(t)\,\|I_\text{d} - I_\text{dp}\|_1, \quad 
    \lambda_{dp}(t) = 
    \begin{cases}
        \lambda_\text{dp}, & t \leq T_\text{dp}, \\
        0, & t > T_\text{dp},
    \end{cases}
\end{equation}
where $I_\text{dp}$ is the diffuse color output image, $\lambda_\text{dp}$ the loss weight of the diffuse prior term and $T_\text{dp}$ its cutoff iteration.


\subsection{Composite Objective}

Our objective loss function follows 2DGS \cite{huang20242dgs} with an RGB reconstruction loss and a D-SSIM loss in sRGB space, 
\begin{equation}
    \mathcal{L}_c = (1-\lambda) \mathcal{L}_1 + \lambda \mathcal{L}_\text{D-SSIM},
\end{equation}
where $\lambda$ is the balancing weight and is set to 0.2 by default. We also use the cosine distance loss $\mathcal{L}_\text{n}$ that enforces consistency between the predicted normals $\hat{N}$ and the normals derived from the depth gradients $\hat{N}_D$. Finally, we add the normal prior loss $\mathcal{L}_\text{np}$, the diffuse prior loss $\mathcal{L}_\text{dp}$, and the white-light regularization loss $\mathcal{L}_\text{light}$ used in \cite{liu2023nero,tong2025gs2dgs,R3DG2024}. The total objective loss function is defined as
\begin{equation}
    \mathcal{L} = \mathcal{L}_\text{c} +
    \sum_{i \in \mathcal{I}} \lambda_i \mathcal{L}_i,
    \quad 
    \mathcal{I} = \{\text{n}, \text{np}, \text{dp}, \text{light}\}.
\end{equation}
where $\lambda_\text{n}$, $\lambda_\text{np}$, $\lambda_\text{dp}$, $\lambda_\text{light}$ are the corresponding loss weights for the utilized losses and regularizers.

\section{Results} \label{sec:results}
\subsection{Implementation Details} \label{sec:results:implementation}
We train synthetic scenes for 50k iterations and real scenes for 20k iterations on an NVIDIA GeForce RTX 4090. Deferred shading is used throughout training. The indirect lighting stage begins at 20k iterations for synthetic scenes and 10k for real scenes. After this point, we extract an object mesh every 3k iterations to enable visibility estimation via ray tracing. Learning rates are set to 0.005 for $F_0$ and roughness, 0.0075 for albedo, and 0.01 for the environment lighting cubemap, each decayed by a factor of 0.1. All other trainable attributes follow the settings of 2DGS \cite{huang20242dgs}. Loss weights are set to $\lambda=0.2$, $\lambda_\text{n}=0.05$, $\lambda_\text{np}=0.01$, $\lambda_\text{dp}=0.05$, and $\lambda_\text{light}=0.001$. The cutoff time for the diffuse prior is set as $T_\text{dp}=15k$, and the prior is disabled when it deteriorates performance in failure cases of StableDelight \cite{stabledelight2025} e.g.~mirror-like objects such as teapot from the Glossy Synthetic dataset \cite{liu2023nero}. For PU, we start with a face resolution $\text{Res}^\text{init}=64$ and progressively upsample the cubemap faces every $15$k iterations until reaching $\text{Res}^\text{final}=512$. The cubemap is prefiltered in every iteration until the base resolution of $\text{Res}^\text{min}=16$ for approximating different levels of roughness from $r^\text{min}=0.02$ to $r^\text{max}=0.5$.

\subsection{Comparison} \label{sec:results:comparison}
We evaluate our methodology against state-of-the-art methods on the tasks of relighting, geometry reconstruction, lighting recovery, and novel view synthesis.

\begin{figure}
    \centering
    \includegraphics[width=\linewidth]{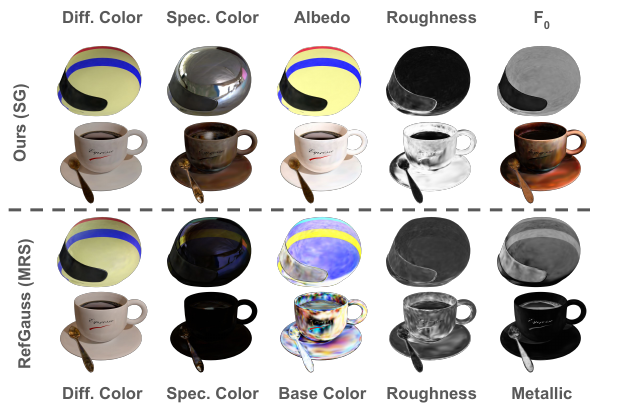}
    \caption{Our method recovers more plausible materials, better suited to downstream tasks like relighting.}
    \label{fig:results:materials}
\end{figure}

\begin{figure}[ht]
    \centering
    \includegraphics[width=\linewidth]{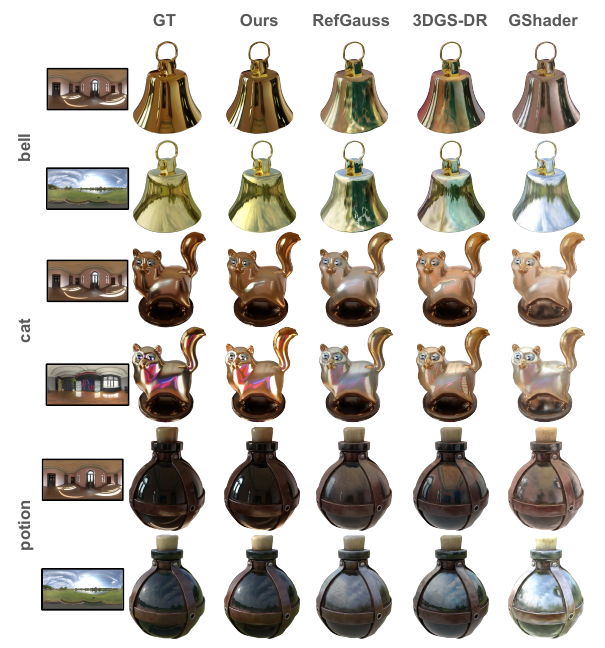}
    \caption{Relighting comparison on three scenes in the Glossy Synthetic \cite{liu2023nero} dataset, with the used environment map on the left.}
    \label{fig:results:relighting}
\end{figure}

\noindent \textbf{Relighting.} \cref{tab:results:relighting} and \cref{fig:results:relighting} present our relighting results. Relighting is performed using an HDR environment map unseen during training, converted from latlong format into a cubemap, then prefiltered across mip levels to approximate varying roughness levels. Our method directly uses the raw HDR radiance values, while other methods apply tone mapping as specified in their pipelines. We do not apply any rescaling based on GT relighted images, or any other post-processing step and report the raw relighted images generated, solely based on recovered material properties and the novel HDR environment map. Both quantitative and qualitative evaluations show that our approach achieves superior relighting performance on complex glossy objects. We also compare results across the three material parameterizations: SG, MR, and MRS, discussed in \cref{sec:methodology:brdf}. Our method with the SG parameterization consistently outperforms the baselines and variants and achieves improved relighting fidelity thanks to more plausible material estimates.  \cref{fig:results:materials} shows the recovered material properties and diffuse/specular outputs for our method vs MRS-based Ref-Gaussian \cite{yao2025refgaussian}. The latter tends to overestimate metallic as can be seen on the visor and stripes of the helmet and the surface of the coffee.

\begin{table}[ht]
\setlength{\tabcolsep}{1.6pt}
\centering
\caption{Relighting evaluation averaged over three previously unseen environment maps (corridor, golf, neon) on the Glossy Synthetic \cite{liu2023nero} scenes. Results for R3DG \cite{R3DG2024} 
and GS-2DGS \cite{tong2025gs2dgs} 
are based on \cite{tong2025gs2dgs}. The rest are obtained through the released codebases. We highlight the \firsttxt{first}, \secondtxt{second}, and \thirdtxt{third} best results.}
\vspace{-0.2cm}
\label{tab:results:relighting}
\renewcommand{\arraystretch}{1.0}  
\resizebox{\linewidth}{!}{%
\begin{tabular}{lccccccc}
\toprule
\multicolumn{8}{c}{\textbf{Relighting}} \\
\midrule
\textbf{Method / Scene} & \textbf{bell} & \textbf{cat} & \textbf{luyu} & \textbf{potion} & \textbf{tbell} & \textbf{teapot} & \textbf{\textit{avg}} \\
\midrule
\multicolumn{8}{c}{\textbf{PSNR} $\uparrow$}\\
\midrule
GShader \cite{jiang2024gshader} & 20.34 & 15.92 & 15.83 & 14.04 & 18.47 & 19.25 & 17.31 \\
R3DG \cite{R3DG2024} & 18.71 & 20.22 & 20.30 & 19.81 & 16.50 & 17.22 & 18.79\\
GS-2DGS \cite{tong2025gs2dgs} & 18.95 & \third{21.78} & 18.82 & 17.88 & 17.55 & 18.96 & 18.99\\
3DGS-DR \cite{ye2024gsdr} & \third{20.59} & 20.36 & \third{21.34} & 20.01 & 19.08 & 19.94 & 20.22 \\
RefGauss \cite{yao2025refgaussian} & 19.87 & 21.58 & 20.45 & 20.20 & \third{21.35} & \second{21.45} & \third{20.82} \\
Ours (MR) & 17.03 & 21.73 & 21.29 & \first{27.57} & 15.81 &	18.62 & 20.34 \\
Ours (MRS) & \second{21.66} & \second{26.14} & \second{21.80} & \second{27.35} & \second{22.94} & \third{20.82} & \second{23.45} \\
Ours (SG) &	\first{25.40} & \first{26.82} & \first{22.73} & \third{27.13} & \first{23.77} & \first{22.72} & \first{24.76}\\
\midrule
\multicolumn{8}{c}{\textbf{SSIM} $\uparrow$}\\
\midrule
GShader \cite{jiang2024gshader}  & \third{0.900} & 0.868 & 0.828 & 0.783 & 0.875 & 0.883 & 0.856 \\
R3DG \cite{R3DG2024} & 0.840 & 0.839 & 0.862 & \third{0.903} & 0.819 & 0.798 & 0.818 \\
GS-2DGS \cite{tong2025gs2dgs} & 0.860 & 0.851 & 0.879 & \second{0.919} & 0.824 & 0.824 & 0.866\\
3DGS-DR \cite{ye2024gsdr} & 0.886 & 0.894 & 0.875 & 0.862 & 0.883 & \third{0.906} & 0.884 \\
RefGauss \cite{yao2025refgaussian} & 0.888 & \second{0.919} & 0.877 & 0.870 & 0.912 & 0.894 & \third{0.893} \\
Ours (MR) & 0.866 & 0.851 & \third{0.902} & 0.856 & \second{0.937} & 0.787 & 0.865 \\
Ours (MRS) & \second{0.914} & \third{0.899} & \first{0.937} & 0.894 & \first{0.938} & \second{0.912} & \second{0.904} \\
Ours (SG) & \first{0.942} & \first{0.946} & \first{0.905} & \first{0.939} & \third{0.930} & \first{0.928} & \first{0.932}\\
\midrule
\multicolumn{8}{c}{\textbf{LPIPS} $\downarrow$}\\
\midrule
GShader \cite{jiang2024gshader} & 0.114 & 0.104 & 0.098 & 0.156 & 0.138 & 0.115 & 0.121 \\
R3DG \cite{R3DG2024} &  - & - & - & - & - & - & -\\
GS-2DGS \cite{tong2025gs2dgs} & - &  - &  - &  - &  - &  - & -\\
3DGS-DR \cite{ye2024gsdr} & \third{0.103} & 0.097 & 0.083 & 0.115 & 0.110 & \third{0.080} & 0.098	\\
RefGauss \cite{yao2025refgaussian} & 0.104 & \third{0.070} & \third{0.079} & 0.110 & \third{0.087} & 0.087 & \third{0.090} \\
Ours (MR) & 0.142 & 0.091 & 0.099 & \third{0.079} & 0.133 & 0.099 & 0.107 \\
Ours (MRS) & \second{0.091} & \second{0.068} & \second{0.070} & \second{0.078} & \second{0.080} & \second{0.079} & \second{0.078} \\
Ours (SG) & \first{0.062} & \first{0.064} & \first{0.065} & \first{0.077} & \first{0.074} & \first{0.067} & \first{0.068}\\
\bottomrule
\end{tabular}
} 
\end{table}

\noindent \textbf{Reconstruction and Lighting Recovery.}
\cref{tab:results:reconstruction} reports geometry reconstruction quality on the Shiny Synthetic dataset \cite{verbin2022refnerf}, environment map quality across two synthetic datasets \cite{verbin2022refnerf, liu2023nero}, average training time, and rendering speed. Geometry is evaluated via MAE$^\circ$ of surface normals, and illumination via LPIPS on extracted HDR latlong maps. Our method consistently outperforms Gaussian Splatting baselines in both tasks, with minimal added training cost, showcasing the benefit of our contributions. Qualitative examples of recovered environment maps are shown in \cref{fig:results:envmaps}, with more results in the supplementary material.

\begin{table}[t]
\centering
\setlength{\tabcolsep}{4pt}
\caption{Comparison of reconstruction quality, environment map recovery, and efficiency.}
\vspace{-0.2cm}
\label{tab:results:reconstruction}
\begin{tabular}{l c c c c}
\hline
\multirow{2}{*}{\textbf{Method}} & \textbf{Normals} & \textbf{Envmap} & \textbf{Train} & \textbf{Render}\\
& MAE$^\circ$ $\downarrow$ & LPIPS $\downarrow$ & hours $\downarrow$ & FPS $\uparrow$\\
\hline
ENVIDR \cite{liang2023envidr} & 2.74 & 0.501  & 5.84 & 1\\ 
GShader \cite{jiang2024gshader} & 7.00 & 0.635 & 0.48 & 28\\
3DGS-DR \cite{ye2024gsdr} & 2.62 & 0.584 & \textbf{0.35} & \textbf{251}\\
Ref-GS \cite{zhang2024refgs} & 2.21 & - & 0.64 & 45 \\
RefGauss \cite{yao2025refgaussian} & 2.15 & 0.560 & 0.58 & 122 \\
Ours (SG) & \textbf{1.57} & \textbf{0.462} & 1.01 & 86 \\
\hline
\end{tabular}
\end{table}

\begin{figure}[ht]
    \centering
    \includegraphics[width=\linewidth, trim=0.2cm 0cm 0 0, clip]{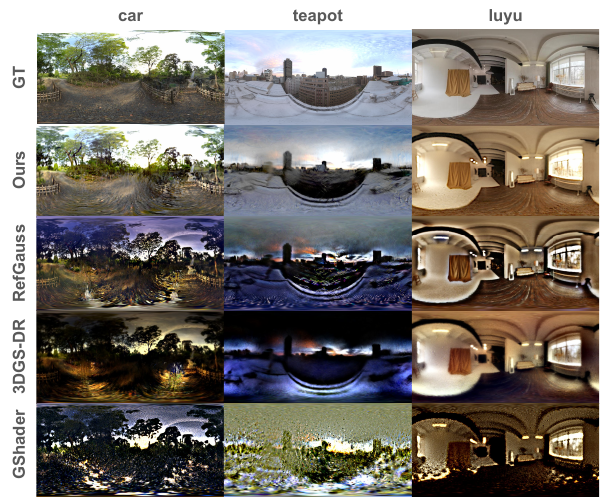}
    \caption{Our method recovers high-quality environment maps with sharper details and fewer artifacts.}
    \vspace{-0.4cm}
    \label{fig:results:envmaps}
\end{figure}

\begin{figure}[ht]
    \centering
    \vspace{-0.4cm}
    \includegraphics[width=\linewidth]{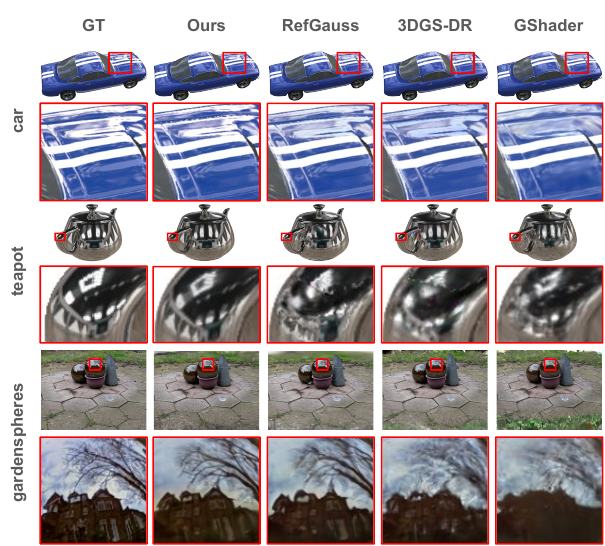}
    \caption{Novel View Synthesis on three scenes from \cite{verbin2022refnerf, liu2023nero}. Our method reconstructs finer details and preserves sharper reflections and specular highlights.}
    \vspace{-0.8cm}
    \label{fig:results:comparisons}
\end{figure}

\noindent \textbf{Novel View Synthesis} \label{sec:results:comparison:nvs}
In \cref{tab:results:comparisons} and \cref{fig:results:comparisons}, we report quantitative and qualitative Novel View Synthesis results against the state-of-the-art. Our method leads on synthetic datasets and is competitive on real scenes, achieving sharper, correctly scaled reflections, as our method retains the high-dynamic-range peaks of the environment lighting that competing methods suppress. Overall, we outperform Gaussian Splatting baselines and train/render faster than implicit baselines. 

\begin{table*}[ht]
\centering
\begin{threeparttable}
\setlength{\tabcolsep}{5pt}
\caption{Comparison of novel view synthesis averaged over scenes within each dataset. The second column denotes which methods support relighting. We highlight the \firsttxt{first}, \secondtxt{second}, and \thirdtxt{third} best results.}
\label{tab:results:comparisons}
\begin{tabular}{l|c|ccc|ccc|ccc}
\toprule
 \multirow{2}{*}{\textbf{Method}} & \multirow{2}{*}{\textbf{Rel.}} & \multicolumn{3}{c|}{\textbf{Shiny Synthetic} \cite{verbin2022refnerf}} & \multicolumn{3}{c|}{\textbf{Glossy Synthetic} \cite{liu2023nero}} & \multicolumn{3}{c}{\textbf{Shiny Real} \cite{verbin2022refnerf}} \\
& & \textbf{PSNR} $\uparrow$ & \textbf{SSIM} $\uparrow$ & \textbf{LPIPS} $\downarrow$ & \textbf{PSNR} $\uparrow$ & \textbf{SSIM} $\uparrow$ & \textbf{LPIPS} $\downarrow$ & \textbf{PSNR} $\uparrow$ & \textbf{SSIM} $\uparrow$ & \textbf{LPIPS} $\downarrow$ \\
\midrule
Ref-NeRF \cite{verbin2022refnerf} & \xmark & 33.13 & 0.961 & 0.080 & 27.50 & 0.927 & 0.100 & 23.62 & 0.646 & 0.239 \\
3DGS~\cite{kerbl20233dgs} & \xmark & 30.36 & 0.947 & 0.084 & 26.50 & 0.916 & 0.092 & 23.85 & 0.660 & \third{0.230}\\
GShader~\cite{jiang2024gshader} & \cmark & 31.97 & 0.958 & 0.067 & 27.54 & 0.922 & 0.087 & 23.46 & 0.647 & 0.257 \\
ENVIDR~\cite{liang2023envidr} & \cmark & 33.46 & \first{0.979} & \first{0.046} & 29.58 & 0.952 & 0.057 & 23.00 & 0.606 & 0.332 \\
3DGS-DR~\cite{ye2024gsdr} & \cmark & 34.09 & 0.971 & 0.057 & 30.22 & 0.953 & 0.061 & 23.99 & \third{0.664} & \second{0.229}\\
Ref-GS~\cite{zhang2024refgs} & \xmark & 34.80 & 0.973 & \third{0.056} & 30.58 & 0.957 & 0.058 & \second{24.44} & 0.657 & \first{0.225}\\
RefGauss~\cite{yao2025refgaussian}\tnote{1}  & \cmark & \second{35.04} & 0.970 & \third{0.056} & \second{30.92} & \second{0.964} & \second{0.047} & 23.98 & 0.663 & 0.278 \\
Ours & \cmark & \first{35.50} & \second{0.978} & \second{0.049} & \first{31.22} & \first{0.966} & \first{0.043} & \third{24.35} & \second{0.679} & 0.259 \\
\bottomrule
\end{tabular}
\begin{tablenotes}
\item[1] \small{RefGauss~\cite{yao2025refgaussian} reported higher scores on Shiny Real, but we were unable to reproduce them and instead report the obtained results.}
\end{tablenotes}
\end{threeparttable}
\end{table*}
\begin{table*}[ht]
\centering
\setlength{\tabcolsep}{8pt}
\caption{Ablation study on the Glossy Synthetic \cite{verbin2022refnerf} dataset comparing relighting, novel view synthesis quality, and training time. PU stands for Progressive Upsampling, Priors denotes the normal and diffuse priors, and NOC stands for Negative-only Clipping.}
\label{tab:results:ablation}
\vspace{-0.2cm}
\begin{tabular}{c c c | c | c c c | c c c | c}
\toprule
\multicolumn{3}{c|}{Components} & \multirow{2}{*}{$\text{Res}^\text{final}$} & \multicolumn{3}{c|}{Relighting}  & \multicolumn{3}{c|}{Novel View Synthesis} & Train Time \\
PU & Priors & NOC & & PSNR\(\uparrow\) & SSIM \(\uparrow\) & LPIPS\(\downarrow\) & PSNR\(\uparrow\) & SSIM \(\uparrow\) & LPIPS\(\downarrow\) & Minutes \(\downarrow\)\\
\hline 
\xmark & \xmark & \xmark & \multirow{4}{*}{512} & 20.99 & 0.912 & 0.075 & 30.52 & 0.964 & 0.046 & 112 \\
\cmark & \xmark  & \xmark & & 21.31 & 0.910 & 0.075 & 30.54 & 0.963 & 0.046 & 61 \\
\cmark & \cmark & \xmark & & 21.00 & 0.913 & 0.077  & 30.80 & 0.965 & 0.044 & 61 \\
\cmark & \cmark & \cmark & & \textbf{24.76} & \textbf{0.932} & \textbf{0.068} & \textbf{31.22} & \textbf{0.966} & \textbf{0.043} & 61 \\
\midrule
\cmark & \cmark & \cmark & 128 & 24.39 & 0.926 & 0.072 & 30.94 & 0.964 & 0.047 & \textbf{40} \\
\bottomrule
\end{tabular}
\end{table*}

\subsection{Ablation Study} \label{sec:results:ablation}

We ablate progressive upsampling (PU), diffuse and normal priors (Priors), and negative-only clipping (NOC) on scenes from the Glossy Synthetic dataset \cite{liu2023nero}. As shown in \cref{tab:results:ablation}, PU cuts training time from 112 to 61 minutes with no NVS drop and a small relighting gain, showing that a coarse-to-fine envmap schedule is beneficial. Priors improve NVS by disambiguating geometry and material properties, yielding cleaner normals and a more stable diffuse/specular split. Relighting changes little with Priors alone because lighting remains clipped and specular cues are bounded. Combining Priors with NOC dramatically improves relighting and also boosts NVS, as Priors reduce geometry-material-lighting ambiguity and NOC preserves the HDR peaks needed for accurate specular IBL.  A secondary benefit of PU is that it makes high-resolution envmaps feasible at reasonable training time. As the last row shows, our final model already performs well at an envmap resolution of 128, and increasing the resolution to 512 yields only a modest lift in relighting and NVS, so the gain is incremental. Our relighting improvements primarily stem from the SG parameterization, PU, Priors, and NOC rather than envmap resolution. Even at low resolution, we are on par with Ref-Gaussian~\cite{yao2025refgaussian} on novel view synthesis, but strongly outperform it on relighting at comparable training time.

\section{Conclusion} \label{sec:conclusion}
We presented an inverse rendering method built on top of 2DGS with pixel-level deferred shading. We proposed to use a Spec-Gloss material parameterization of a microfacet BRDF for improving identifiability in the inverse rendering task. Shading follows the split-sum IBL paradigm with a differentiable cube-mipmap environment lighting, progressively upsampled (PU) for efficiency, with negative-only clipping to preserve HDR lighting peaks. Normal and diffuse priors (Priors) guide early optimization and alleviate the inherent ambiguity of inverse rendering. Experimental results demonstrate state-of-the-art relighting, environment map recovery, and reconstruction while being competitive on novel view synthesis. Ablations confirm that PU reduces training time and that Priors plus NOC provide the largest boost in performance. At the same time, PU facilitates the use of larger resolution cubemaps at minimal cost.
\section*{Acknowledgments}
\begin{sloppypar}
This work was supported by the Flanders AI Research Program and the KU Leuven Methusalem project "Lifelines".
The resources and services used in this work were provided by the VSC (Flemish Supercomputer Center), funded by the Research Foundation - Flanders (FWO) and the Flemish Government.
\end{sloppypar}

{
    \small
    \bibliographystyle{ieeenat_fullname}
    \bibliography{main}
}
\clearpage
\setcounter{page}{1}
\maketitlesupplementary

\section{Indirect Illumination} \label{sec:indirect}
We adopt the indirect lighting formulation of Ref-Gaussian~\cite{yao2025refgaussian}. The specular IBL is split into direct and indirect components, modulated by visibility $V \in \{0,1\}$ along the reflected direction $\boldsymbol{\omega}_r=2(\boldsymbol{\omega}_o \cdot \mathbf{n})\mathbf{n}-\boldsymbol{\omega}_o$. The visibility is computed via bounding volume hierarchy (BVH) for accelerated ray tracing on an extracted TSDF mesh, which is updated periodically (i.e. every 3k iterations) for efficiency. The visible part uses the standard prefiltered environment lookup, while the occluded part is modeled per Gaussian with low-order spherical harmonics $S_i(\cdot)$ evaluated at the reflection direction $\boldsymbol{\omega}_r$ and alpha-blended in screen space:
\begin{equation}
L_{\text{ind}}(\boldsymbol{\omega}_r)=\sum_{i=1}^N l_\text{i}(\boldsymbol{\omega}_r)\,\alpha_i\ \prod_{j=1}^{i-1}(1-\alpha_j),
\end{equation}
where $l_\text{i}(\boldsymbol{\omega}_r)=S_i(\boldsymbol{\omega}_r)$. This formulation captures inter-reflections while retaining real-time performance.

\section{Material Parameterizations in the Literature}
In the following equations, we present the computation of the outgoing color for the various material parameterizations. The Disney BRDF with the metallic-roughness parameterization estimates $L_o$ as 
\begin{equation}\label{eq:disney}
    L_o^\text{MR} = (1-m) b L_d + (F_0(m, b) \beta_1 +\beta_2) L_s,
\end{equation}
which entangles the base color $b$ to the metallic $m$ through $F_0$. GaussianShader \cite{jiang2024gshader} simplifies the rendering equation to solve the entanglement issue by learning $F_0$ and using it for energy conservation on the diffuse appearance as well, departing, however, from principled models in literature or industry. The simplified outgoing radiance equation is defined as:
\begin{equation}\label{eq:simple}
    L_o^\text{Simple} = \sigma(b - ln3) + (1-k_s) L_d + (k_s \beta_1 + \beta_2) L_s,
\end{equation}
where $\sigma$ is the sigmoid function. 
More recently, Ref-Gaussian \cite{yao2025refgaussian} tried a similar approach that solved the entanglement issue by replacing base color $b$ in $F_0$ with a specular tint term $k_s$
\begin{equation}\label{eq:hybrid}
    L_o^\text{MRS} = b + (F_0(m, k_s) \beta_1 +\beta_2) L_s,
\end{equation}
where the Fresnel coefficients $\beta_1$, $\beta_2$ are given from a lookup table indexed with roughness $r$ and the $nv$ product of the surface normal $\mathbf{n}$ and viewing ray direction $\mathbf{v}$. Finally, $L_d$ is the irradiance from the environment, convolved with a diffuse kernel, and $L_s$ is the prefiltered specular environment.
Nevertheless, it faces the same portability issues as GaussianShader, and while both methods perform well on the NVS task, performance deteriorates in relighting.


\section{Additional Results}

\subsection{Scene Editing}
\begin{figure*}[ht]
    \centering
    \includegraphics[width=\linewidth,trim=0cm 0 0cm 0,clip]{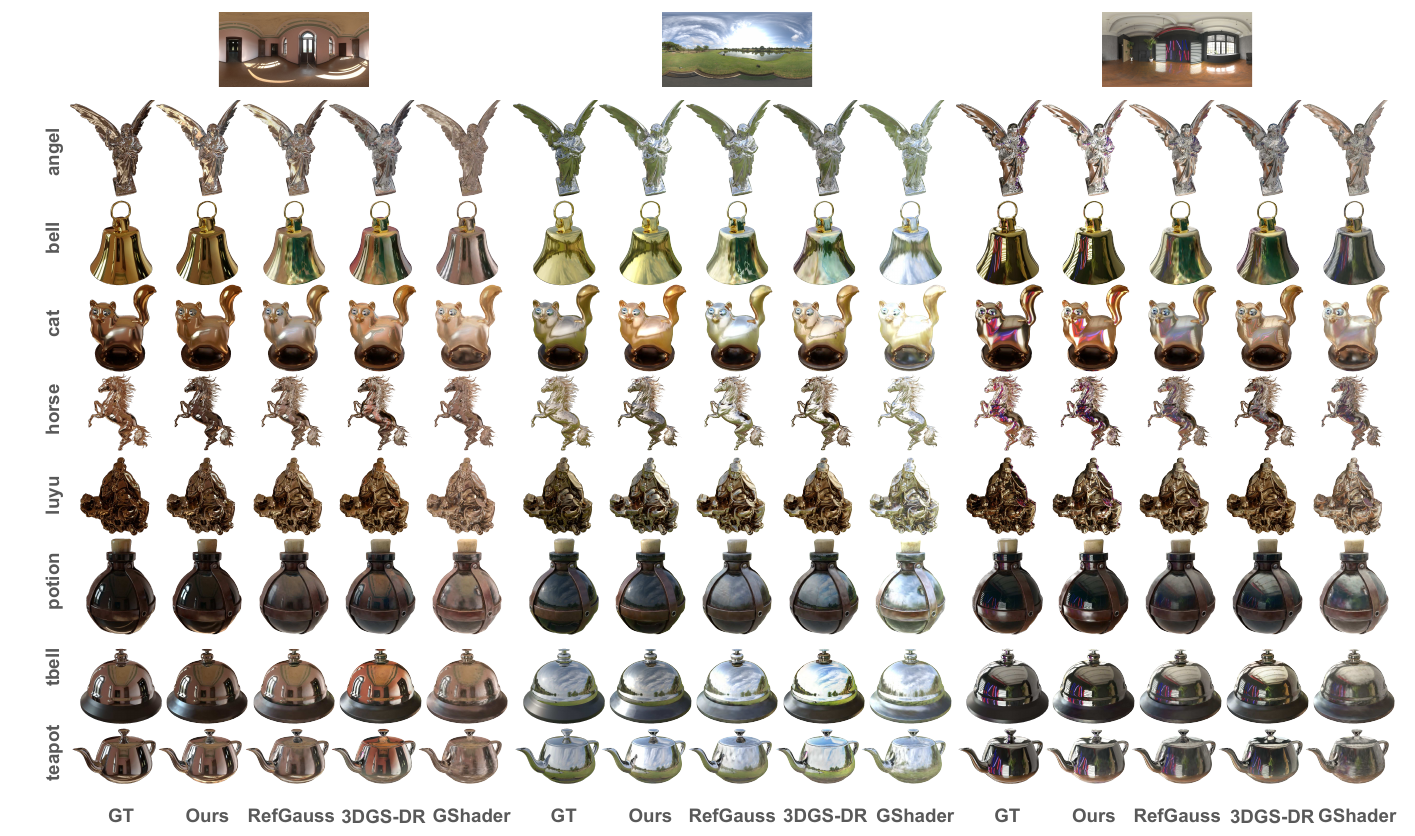}
    \caption{Qualitative relighting comparison under three environment maps (corridor, golf, neon) on the Glossy Synthetic dataset. Our method typically yields more faithful relighting than competing Gaussian Splatting baselines.}
    \label{fig:supplementary:relighting:glossy:comparison}
\end{figure*}

\begin{figure*}[ht]
    \centering
    \includegraphics[width=\linewidth,trim=0cm 0 0cm 0,clip]{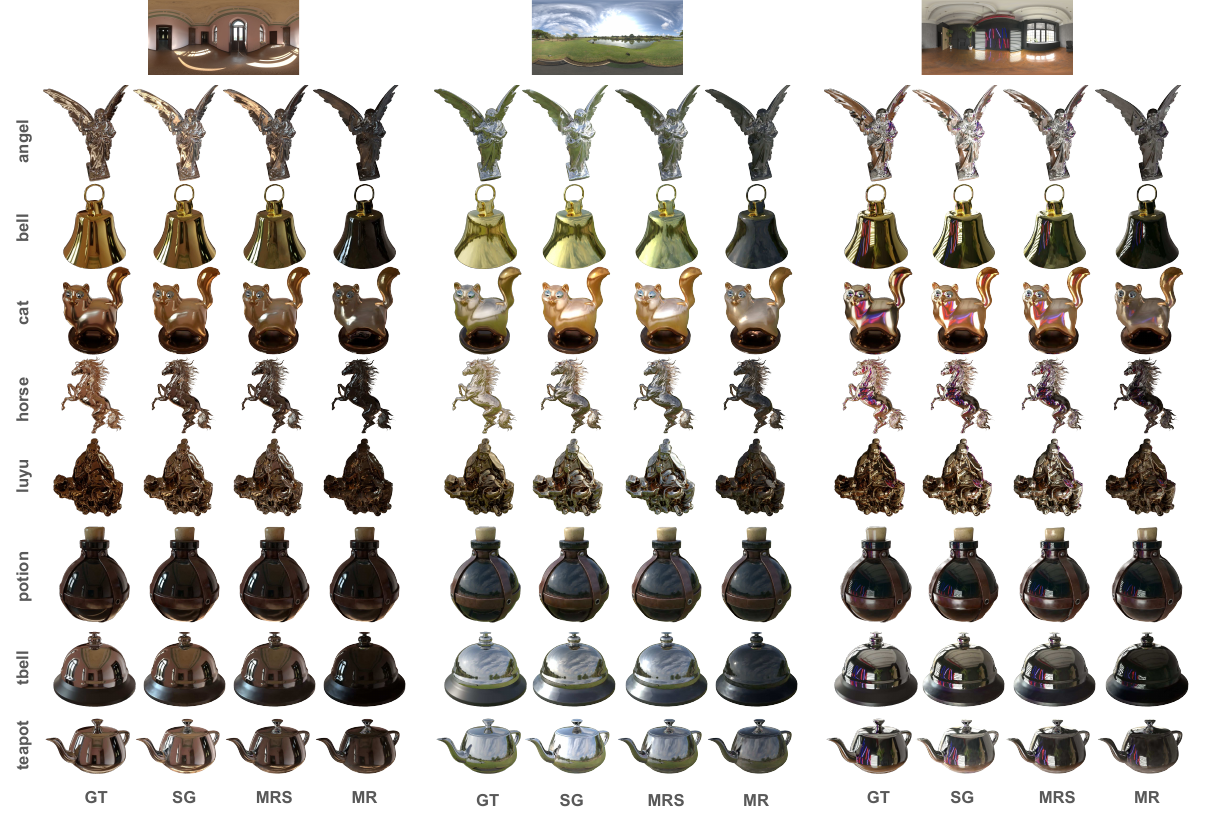}
    \caption{Qualitative relighting comparison under three environment maps (corridor, golf, neon) on the Glossy Synthetic dataset. We compare our method against three alternative material parameterizations, and the results support our choice of the specular–glossiness (SG) parameterization.}
    \label{fig:supplementary:relighting:glossy:brdf}
\end{figure*}

\begin{figure*}
    \centering
    \includegraphics[width=\linewidth]{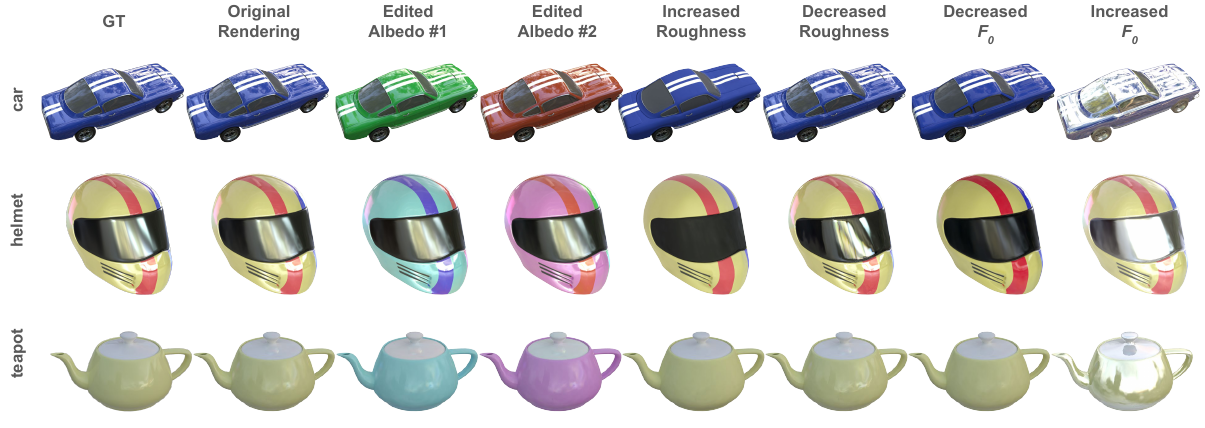}
    \caption{Demonstration of the material-editing capabilities of our method. After training, we modify the reconstructed materials and re-render three scenes from the Shiny Synthetic dataset~\cite{verbin2022refnerf} with increased or decreased albedo, roughness, or $F_0$.}
    \label{fig:supplementary:editing}
\end{figure*}

\cref{tab:supplementary:relighting,fig:supplementary:relighting:glossy:comparison} compare our relighting results with Gaussian Splatting baselines, showing that our method produces more plausible relighting than Ref-Gaussian~\cite{yao2025refgaussian}, 3DGS-DR~\cite{ye2024gsdr}, and GaussianShader~\cite{jiang2024gshader}. \cref{fig:supplementary:relighting:glossy:brdf} compares the three material parameterizations (SG, MR, MRS) and validates our choice, as SG consistently outperforms the MR and MRS variants. \cref{fig:supplementary:editing} further illustrates the material editing capabilities of our method.

\begin{table*}[ht]
\centering
\caption{Per-scene relighting and NVS comparison on the Glossy Synthetic \cite{liu2023nero} scenes. We highlight the \firsttxt{first}, \secondtxt{second}, and \thirdtxt{third} best results.}
\label{tab:supplementary:relighting}
\renewcommand{\arraystretch}{1.0}  
\resizebox{\linewidth}{!}{%
\begin{tabular}{l|ccccccc|ccccccc}
\toprule
 & \multicolumn{7}{c|}{\textbf{Relighting}} 
 & \multicolumn{7}{c}{\textbf{Novel View Synthesis}} \\
 & \textbf{bell} & \textbf{cat} & \textbf{luyu} & \textbf{potion} & \textbf{tbell} & \textbf{teapot} & \textbf{\textit{avg}} & \textbf{bell} & \textbf{cat} & \textbf{luyu} & \textbf{potion} & \textbf{tbell} & \textbf{teapot} & \textbf{\textit{avg}} \\
\midrule
\multicolumn{14}{c}{\textbf{PSNR} $\uparrow$}\\
\midrule
GShader \cite{jiang2024gshader}
    & 20.34 & 15.92 & 15.83 & 14.04 & 18.47 & 19.25 & 17.31
    & 28.07 & 31.81 & 27.18 & 30.09 & 24.48 & 23.58 & 27.55
  \\
3DGS-DR \cite{ye2024gsdr} 
    & \third{20.59} & 20.36 & \third{21.34} & 20.01 & 19.08 & 19.94 & 20.22
    & \third{31.65} & \first{33.86} & {28.71} & {32.79} & {28.94} & 25.36 & {30.14}
  \\
RefGauss \cite{yao2025refgaussian}
    & 19.87 & 21.58 & 20.45 & 20.20 & \third{21.35} & \second{21.45} & \third{20.82}
    & \second{32.86} & {33.01} & \first{30.04} & {33.07} & \third{29.84} & \second{26.68} & \second{30.92}
  \\
Ours (MR)
    & 17.03 & 21.73 & 21.29 & \first{27.57} & 15.81 &	18.62 & 20.34
    & 28.61 & 32.20 & 27.95 & \third{33.31} & 27.59 & 24.05 & 28.95
  \\  
Ours (MRS)
    & \second{21.66} & \second{26.14} & \second{21.80} & \second{27.35} & \second{22.94} & \third{20.82} & \second{23.45}
    & 30.71 & \third{33.11} & \third{29.46} & \second{33.38} & \second{30.01} & \third{25.76} & \third{30.40}
  \\
Ours (SG)
    & \first{25.40} & \first{26.82} & \first{22.73} & \third{27.13} & \first{23.77} & \first{22.72} & \first{24.76}
    & \first{33.40} & \second{33.54} & \second{29.73} & \first{33.57} & \first{30.12} & \first{26.96} & \first{31.22}
  \\
\midrule
\multicolumn{14}{c}{\textbf{SSIM} $\uparrow$}\\
\midrule
GShader \cite{jiang2024gshader}  
  & \third{0.900} & 0.868 & 0.828 & 0.783 & 0.875 & 0.883 & 0.856
  & 0.919 & 0.961 & 0.914 & 0.938 & 0.898 & 0.901 & 0.921
\\
3DGS-DR \cite{ye2024gsdr} & 0.886 & 0.894 & 0.875 & 0.862 & 0.883 & \third{0.906} & 0.884
  & \third{0.962} & \second{0.976} & {0.936} & {0.957} & {0.952} & 0.936 & \third{0.953}
  \\
RefGauss \cite{yao2025refgaussian} 
  & 0.888 & \second{0.919} & 0.877 & 0.870 & 0.912 & 0.894 & \third{0.893}
  & \second{0.969} & \third{0.973} & \second{0.952} & \third{0.963} & \third{0.962} & \second{0.947} & \second{0.961}
  \\
Ours (MR) 
    & 0.866 & 0.851 & \third{0.902} & 0.856 & \second{0.937} & 0.787 & 0.865
    & 0.944 & 0.970 & 0.934 & \second{0.967} & 0.948 & 0.929 & 0.949
  \\  
Ours (MRS) 
    & \second{0.914} & \third{0.899} & \first{0.937} & 0.894 & \first{0.938} & \second{0.912} & \second{0.904}
    & \third{0.962} & \second{0.976} & \third{0.950} & \first{0.968} & \second{0.966} & \third{0.946} & \second{0.961}
  \\
Ours (SG)
    & \first{0.942} & \first{0.946} & \first{0.905} & \first{0.939} & \third{0.930} & \first{0.928} & \first{0.932}
    & \first{0.975} & \first{0.978} & \first{0.953} & \first{0.968} & \first{0.967} & \first{0.956} & \first{0.966}
  \\
\midrule
\multicolumn{14}{c}{\textbf{LPIPS} $\downarrow$}\\
\midrule
GShader \cite{jiang2024gshader}
    & 0.114 & 0.104 & 0.098 & 0.156 & 0.138 & 0.115 & 0.121
    & 0.098 & 0.056 & 0.064 & 0.088 & 0.122 & 0.091 & 0.086
    \\
3DGS-DR \cite{ye2024gsdr}
    & \third{0.103} & 0.097 & 0.083 & 0.115 & 0.110 & \third{0.080} & 0.098
    & 0.064 & \third{0.040} & \third{0.053} & {0.075} & {0.067} & 0.067 & 0.058
    \\
RefGauss \cite{yao2025refgaussian}
    & 0.104 & \third{0.070} & \third{0.079} & 0.110 & \third{0.087} & 0.087 & \third{0.090}
    & \second{0.040} & \third{0.040} & \second{0.043} & \third{0.064} & \third{0.058} & \third{0.058} & \third{0.051}
    \\
Ours (MR)
    & 0.142 & 0.091 & 0.099 & \third{0.079} & 0.133 & 0.099 & 0.107
    & 0.072 & 0.047 & 0.055 & \second{0.054} & 0.072 & 0.069 & 0.062
    \\
Ours (MRS) 
    & \second{0.091} & \second{0.068} & \second{0.070} & \second{0.078} & \second{0.080} & \second{0.079} & \second{0.078}
   & \third{0.045} & \second{0.038} & \second{0.043} & \second{0.054} & \first{0.049} & \second{0.055} & \second{0.047}
    \\
Ours (SG) 
    & \first{0.062} & \first{0.064} & \first{0.065} & \first{0.077} & \first{0.074} & \first{0.067} & \first{0.068} \second
    & \first{0.032} & \first{0.035} & \first{0.042} & \first{0.053} & \second{0.050} & \first{0.046} & \first{0.043}
    \\
\bottomrule
\end{tabular}
} 
\end{table*}


\subsection{Recovered Environment Map Results}
\begin{figure*}[ht]
    \centering
    \includegraphics[width=\linewidth]{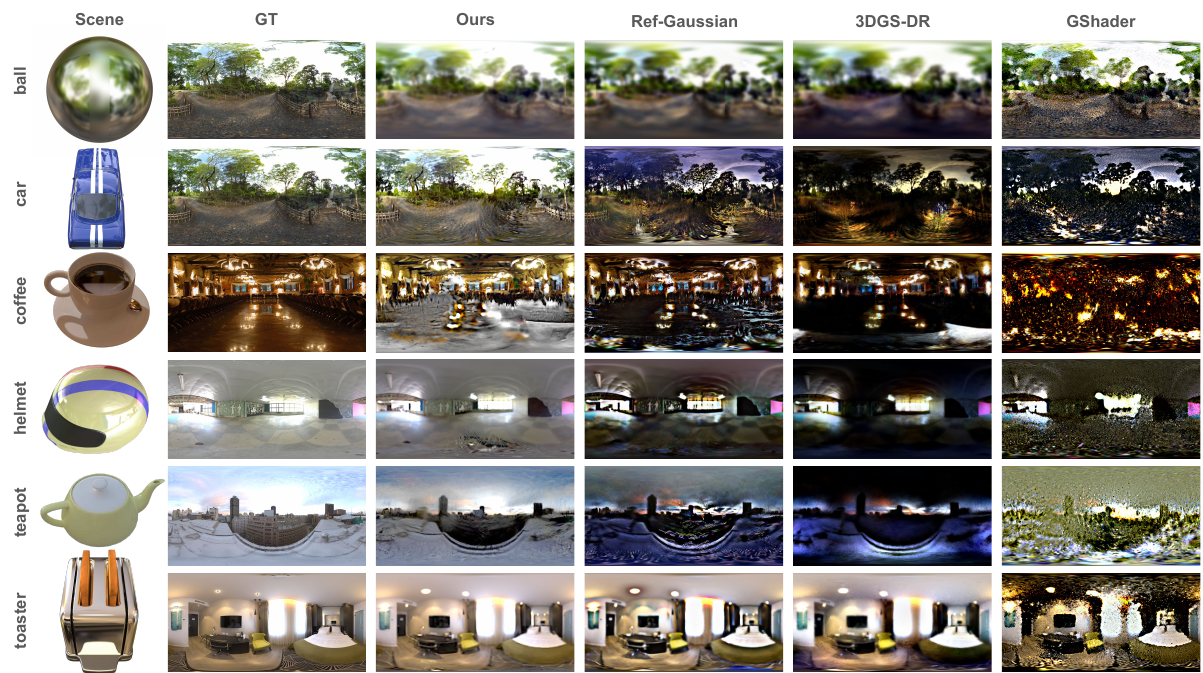}
    \caption{Comparison of recovered environment maps on the scenes of the Shiny Synthetic dataset~\cite{verbin2022refnerf}.}
    \label{fig:supplementary:envmaps:shiny}
\end{figure*}

\begin{figure*}[ht]
    \centering
    \includegraphics[width=\linewidth]{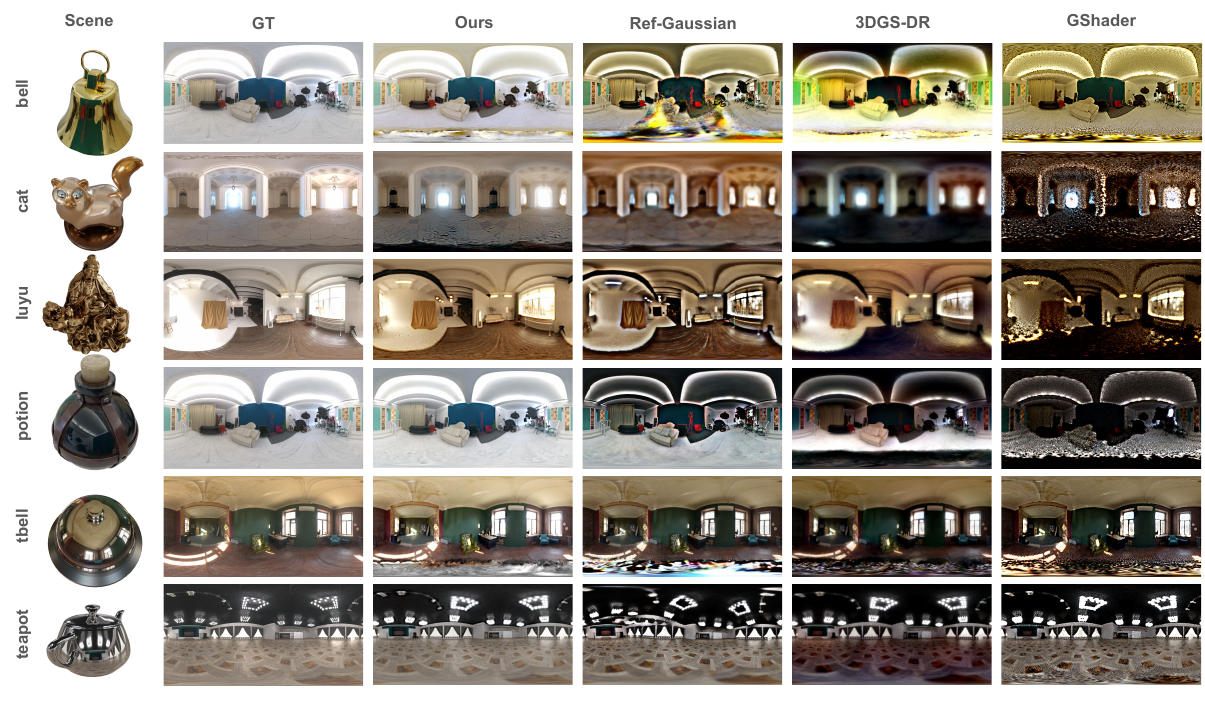}
    \caption{Comparison of recovered environment maps from the scenes of the Glossy Synthetic dataset~\cite{liu2023nero}.}
    \label{fig:supplementary:envmaps:glossy}
\end{figure*}

\begin{figure*}[ht]
    \centering
    \includegraphics[width=\linewidth]{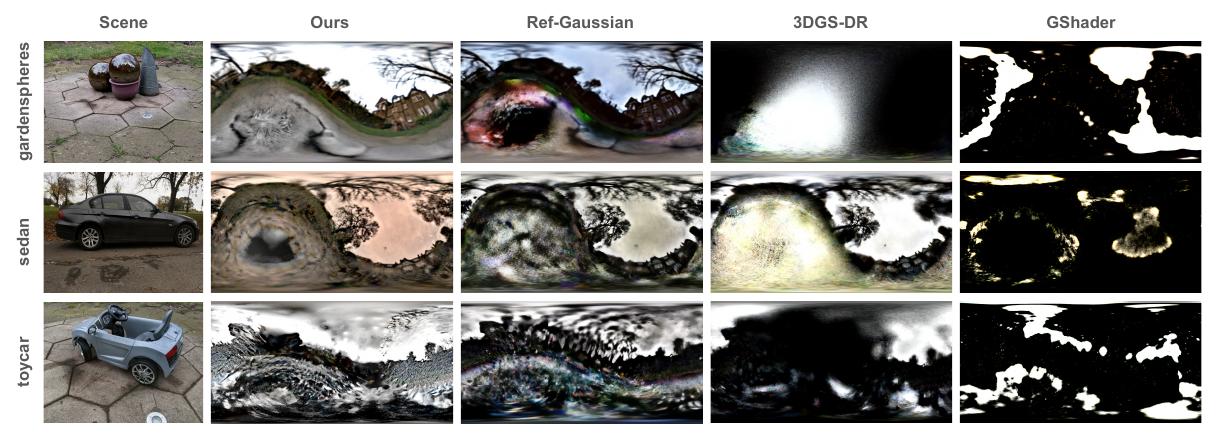}
    \caption{Comparison of recovered environment maps on the scenes of the Shiny Real dataset~\cite{verbin2022refnerf}.}
    \label{fig:supplementary:envmaps:real}
\end{figure*}

Figures \ref{fig:supplementary:envmaps:shiny}-\ref{fig:supplementary:envmaps:real} present additional results on environment map recovery for the Shiny Synthetic \cite{verbin2022refnerf}, Glossy Synthetic \cite{liu2023nero}, and Shiny Real \cite{verbin2022refnerf} scenes.


\subsection{Reconstruction Results}
\begin{figure*}[ht]
    \centering
    \includegraphics[width=\linewidth]{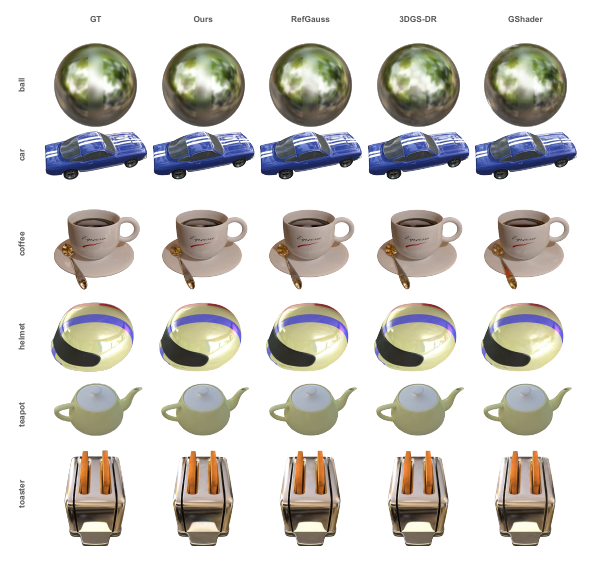}
    \caption{Comparison of rendering quality on the Shiny Synthetic dataset \cite{verbin2022refnerf}.}
    \label{fig:supplementary:comparisons:shiny}
\end{figure*}

\begin{figure*}[ht]
    \centering
    \includegraphics[width=\linewidth]{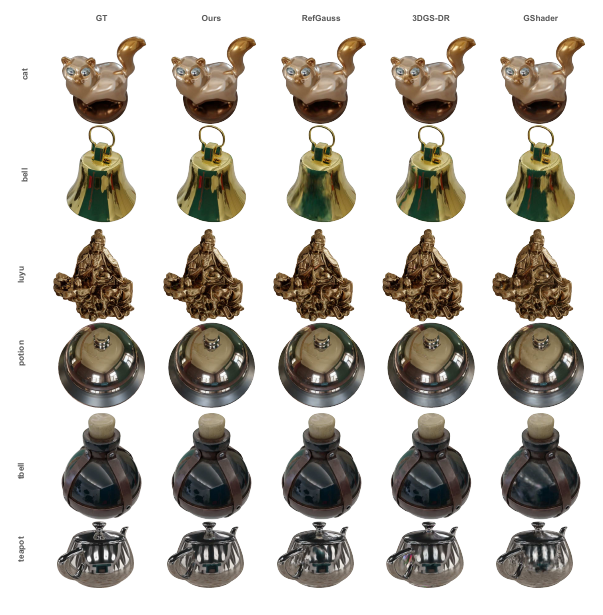}
    \caption{Comparison of rendering quality on the Glossy Synthetic dataset \cite{liu2023nero}.}
    \label{fig:supplementary:comparisons:glossy}
\end{figure*}

\begin{figure*}[ht]
    \centering
    \includegraphics[width=\linewidth]{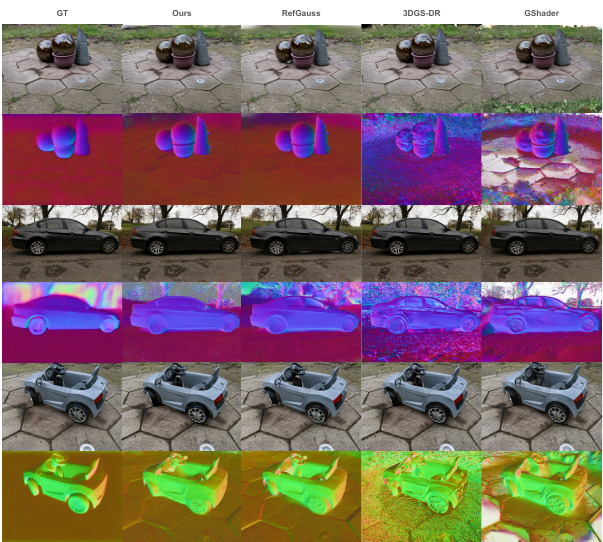}
    \caption{Comparison of rendering quality on the Shiny Real dataset \cite{verbin2022refnerf}. In the leftmost column, we show the GT RGB test images and the pseudo-GT surface normals, generated by StableNormal \cite{ye2024stablenormal}, and used to supervise the predicted normals.}
    \label{fig:supplementary:comparisons:real}
\end{figure*}

In Figures~\ref{fig:supplementary:comparisons:shiny} - \ref{fig:supplementary:comparisons:real}, we present qualitative results for all examined scenes. The corresponding quantitative results are presented in \cref{tab:supplementary:comparisons}. We observe that our method is competitive and often outperforms Gaussian splatting and implicit baselines. The gap in performance in real scenes is mostly due to the limitation of 2DGS in representing dense, thin structures and sparsely seen background features.

\begin{table*}[ht]
\setlength{\tabcolsep}{1.6pt}
\centering
\caption{Quantitative comparison of novel view synthesis averaged over scenes within each dataset. Methods labeled with \cmark in the Rel. column support relighting while those labeled with \xmark do not. For RefGauss \cite{yao2025refgaussian}, although higher scores were reported on Shiny Real  \cite{verbin2022refnerf}, we were unable to reproduce them. We therefore report results obtained using the authors’ released code. We highlight the \firsttxt{first}, \secondtxt{second}, and \thirdtxt{third} best results.}
\label{tab:supplementary:comparisons}
\renewcommand{\arraystretch}{1.0}  
\resizebox{\linewidth}{!}{%
\begin{tabular}{l|c|ccccccc|ccccccc|cccc}
\toprule
\multirow{2}{*}{\textbf{Method}} & \multirow{2}{*}{\textbf{Rel.}} & \multicolumn{7}{c|}{\textbf{Shiny Synthetic \cite{verbin2022refnerf}}} 
 & \multicolumn{7}{c|}{\textbf{Glossy Synthetic \cite{liu2023nero}}}
 & \multicolumn{4}{c}{\textbf{Shiny Real \cite{verbin2022refnerf}}}\\
 & & \textbf{ball} & \textbf{car} & \textbf{coffee} & \textbf{helmet} & \textbf{teapot} & \textbf{toaster} & \textbf{\textit{avg}}
& \textbf{bell} & \textbf{cat} & \textbf{luyu} & \textbf{potion} & \textbf{tbell} & \textbf{teapot} & \textbf{\textit{avg}}
& \textbf{garden} & \textbf{sedan} & \textbf{toycar} & \textbf{\textit{avg}}\\
\midrule
\multicolumn{19}{c}{\textbf{PSNR} $\uparrow$}\\
\midrule
Ref-NeRF & \xmark
  & 33.16 & 30.44 & 33.99 & 29.94 & 45.12 & 26.12 & 33.13  
  & 30.02 & 29.76 & 25.42 & 30.11 & 26.91 & 22.77 & 27.50
  & 22.01 & 25.21 & 23.65 & 23.62 
  \\
3DGS & \xmark
  & 27.65 & 27.26 & 32.30 & 28.22 & 45.71 & 20.99 & 30.36
  & 25.11 & 31.36 & 26.97 & 30.16 & 23.88 & 21.51 & 26.50
  & 21.75 & 26.03 & 23.78 & 23.85
  \\
GShader & \cmark
  & 30.99 & 27.96 & 32.39 & 28.32 & 45.86 & 26.28 & 31.97 
  & 28.07 & 31.81 & 27.18 & 30.09 & 24.48 & 23.58 & 27.54
  & 21.74 & 24.89 & 23.76 & 23.46
  \\
ENVIDR & \cmark    
  & \first{41.02} & 27.81 & 30.57 & \third{32.71} & 42.62 & 26.03 & 33.46  
  & 30.88 & 31.04 & 28.03 & 32.11 & 28.64 & \second{26.77} & 29.58 
  & 21.47 & 24.61 & 22.92 & 23.00
  \\
3DGS-DR & \cmark
  & 33.66 & 30.39 & \second{34.65} & 31.69 & \second{47.12} & 27.02 & 34.09
  & 31.65 & \first{33.86} & 28.71 & \third{32.79} & 28.94 & 25.36 & 30.22
  & 21.82 & \second{26.32} & 23.83 & \third{23.99}
  \\
Ref-GS & \xmark   
  & 36.10 & \third{30.94} & 34.38 & \second{33.40} & 46.69 & \third{27.28} & \third{34.80}
  & \third{31.70} & \third{33.15} & \third{29.46} & 32.64 & \second{30.08} & 26.47 & \third{30.58}       
  & \third{22.48} & \first{26.63} & \first{24.20} & \first{24.44}
  \\
RefGauss & \cmark
  & \third{37.01} & \second{31.04} & \third{34.63} & 32.32 & \first{47.16} & \first{28.05} & \second{35.04}
  & \second{32.86} & 33.01 & \first{30.04} & \second{33.07} & \third{29.84} & \third{26.68} & \second{30.92}
  & \first{22.79} & 25.13 & \third{24.01} & 23.98
  \\
Ours & \cmark
  & \second{38.14} & \first{31.92} & \first{34.72} & \first{33.72} & \third{46.77} & \second{27.69} & \first{35.50}
  & \first{33.40} & \second{33.54} & \second{29.73} & \first{33.57} & \first{30.12} & \first{26.96} & \first{31.22}
  & \second{22.69} & \third{26.17} & \second{24.19} & \second{24.35}
  \\
\midrule
\multicolumn{19}{c}{\textbf{SSIM} $\uparrow$}\\
\midrule
Ref-NeRF & \xmark
  & 0.971 & 0.950 & 0.972 & 0.954 & \third{0.995} & 0.921 & 0.961                 
  & 0.941 & 0.944 & 0.901 & 0.933 & 0.947 & 0.897 & 0.927
  & \third{0.584} & 0.720 & 0.633 & 0.646 
  \\
3DGS & \xmark        
  & 0.937 & 0.931 & 0.972 & 0.951 & \second{0.996} & 0.894 & 0.947  
  & 0.892 & 0.959 & 0.916 & 0.938 & 0.908 & 0.881 & 0.916
  & 0.571 & \third{0.771} & 0.637 & 0.660  
  \\                                                               
GShader & \cmark     
  & 0.966 & 0.932 & 0.971 & 0.951 & \second{0.996} & 0.929 & 0.958 
  & 0.919 & 0.961 & 0.914 & 0.938 & 0.898 & 0.901 & 0.922
  & 0.576 & 0.728 & 0.637 & 0.647
\\
ENVIDR & \cmark
  & \first{0.997} & 0.943 & 0.962 & \first{0.987} & \third{0.995} & \first{0.990} & \first{0.979}
  & 0.954 & 0.965 & 0.931 & \third{0.960} & 0.947 & \first{0.957} & 0.952   
  & 0.561 & 0.707 & 0.549 & 0.606
  \\
3DGS-DR & \cmark
  & 0.979 & \third{0.962} & \second{0.976} & 0.971 & \first{0.997} & 0.943 & 0.971   
  & 0.962 & \second{0.976} & 0.936 & 0.957 & 0.952 & 0.936 & 0.953
  & 0.581 & \second{0.773} & 0.639 & \second{0.664}
  \\
Ref-GS & \xmark                                                                  
  & \third{0.981} & 0.961 & \third{0.973} & \third{0.975} & \first{0.997} & \second{0.950} & \third{0.973}
  & \third{0.965} & \third{0.973} & \third{0.946} & 0.957 & \third{0.956} & 0.944 & \third{0.957}
  & 0.507 & \first{0.783} & \first{0.682} & 0.657  
  \\
RefGauss & \cmark
  & \third{0.981} & \second{0.964} & \second{0.976} & 0.959 & \first{0.997} & 0.942 & 0.970
  & \second{0.969} & \third{0.973} & \second{0.952} & \second{0.963} & \second{0.962} & \third{0.947} & \second{0.964}
  & \first{0.616} & 0.731 & \third{0.642} & \third{0.663}
  \\
Ours & \cmark
    & \second{0.989} & \first{0.974} & \first{0.977} & \second{0.978} & \first{0.997} & \third{0.949} & \second{0.978}
    & \first{0.975} & \first{0.978} & \first{0.953} & \first{0.968} & \first{0.967} & \second{0.956} & \first{0.966} 
    & \second{0.615} & 0.761 & \second{0.661} & \first{0.679}
  \\
\midrule
\multicolumn{19}{c}{\textbf{LPIPS} $\downarrow$}\\
\midrule
Ref-NeRF & \xmark
  & 0.166 & 0.050 & 0.082 & 0.086 & 0.012 & 0.083 & 0.080 
  & 0.102 & 0.104 & 0.098 & 0.084 & 0.114 & 0.098 & 0.100
  & 0.251 & 0.234 & \first{0.231} & 0.239 
  \\
3DGS & \xmark             
  & 0.162 & 0.047 & \third{0.079} & 0.081 & 0.008 & 0.125 & 0.084         
  & 0.104 & 0.062 & 0.064 & 0.093 & 0.125 & 0.102  & 0.092
  & \third{0.248} & \second{0.206} & \third{0.237} & \third{0.230}
  \\
GShader & \cmark    
  & 0.121 & 0.044 & \second{0.078} & 0.074 & \third{0.007} & 0.079 & 0.067   
  & 0.098 & 0.056 & 0.064 & 0.088 & 0.122 & 0.091 & 0.087
  & 0.274 & 0.259 & 0.239 & 0.257
  \\
ENVIDR & \cmark    
  & \first{0.020} & 0.046 & 0.083 & \first{0.036} & 0.009 & 0.081 & \first{0.046}
  & 0.054 & 0.049 & 0.059 & \third{0.072} & 0.069 & \first{0.041} & \third{0.057}
  & 0.263 & 0.387 & 0.345 & 0.332
  \\
3DGS-DR & \cmark
  & \third{0.098} & \second{0.033} & \first{0.076} & 0.049 & \first{0.005} & 0.081 & 0.057        
  & 0.064 & \second{0.040} & 0.053 & 0.075 & \third{0.067} & 0.067 & 0.061    
  & \second{0.247} & \third{0.208} & \first{0.231} & \second{0.229}
  \\
Ref-GS &  \xmark    
  & \third{0.098} & \third{0.034} & 0.082 & 0.045 & \second{0.006} & \first{0.070} & \third{0.056}
  & \third{0.049} & \third{0.041} & \third{0.046} & 0.076 & 0.073 & 0.064 & 0.058
  & \first{0.242} & \first{0.196} & \second{0.236} & \first{0.225}
  \\
RefGauss & \cmark
  & \third{0.098} & \second{0.033} &\first{0.076} & \third{0.050} & \second{0.006} & \third{0.074} & \third{0.056}
  & \second{0.040} & \second{0.040} & \second{0.043} & \second{0.064} & \second{0.058} & \third{0.058} & \second{0.047}
  & 0.278 & 0.277 & 0.279 & 0.278
  \\
Ours & \cmark
  & \second{0.073} & \first{0.027} & \second{0.078} & \second{0.038} & \second{0.006} & \second{0.073} & \second{0.049}
  & \first{0.032} & \first{0.035} & \first{0.042} & \first{0.053} & \first{0.050} & \second{0.046} & \first{0.043}
  & 0.291 & 0.222 & 0.264 & 0.259
  \\
\bottomrule
\end{tabular}
} 
\end{table*}



\subsection{Material Decomposition}
\begin{figure*}
    \centering
    \includegraphics[width=0.9\linewidth]{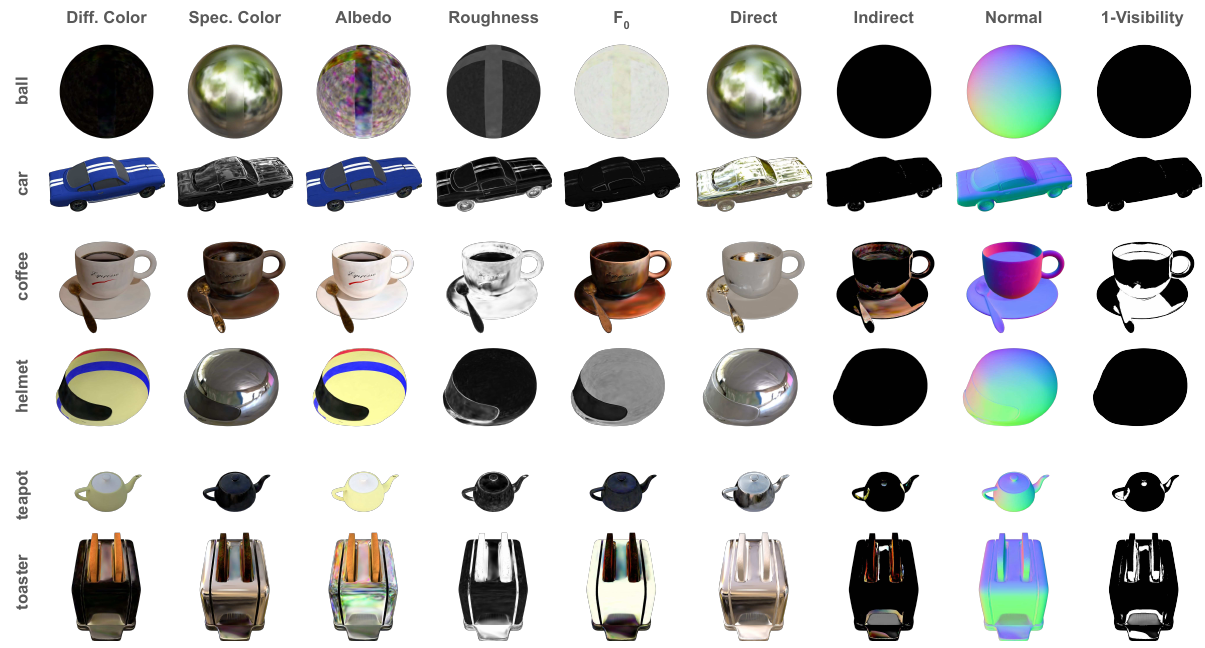}
    \caption{Material decomposition and outputs of our method on the Shiny Synthetic dataset \cite{verbin2022refnerf}}
    \label{fig:supplementary:materials:shiny}
\end{figure*}

\begin{figure*}
    \centering
    \includegraphics[width=0.9\linewidth]{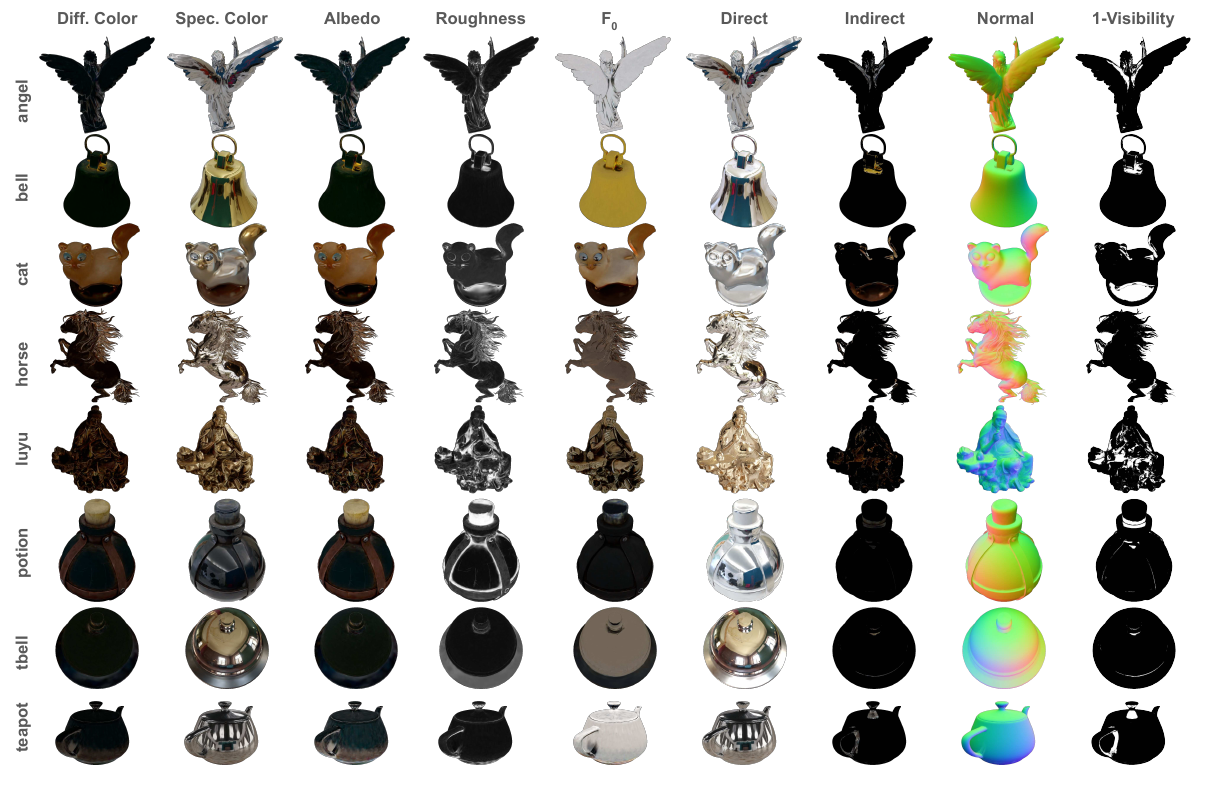}
    \caption{Material decomposition and outputs of our method on the Glossy Synthetic dataset \cite{liu2023nero}}
    \label{fig:supplementary:materials:glossy}
\end{figure*}

In \cref{fig:supplementary:materials:shiny} and \cref{fig:supplementary:materials:glossy}, we present the material decomposition of our method, the diffuse and specular components, the surface normals, as well as visibility, direct and indirect light terms for all examined synthetic scenes.

\subsection{Diffuse Color and Surface Normal Priors}
\begin{figure*}
    \centering
    \includegraphics[width=\linewidth]{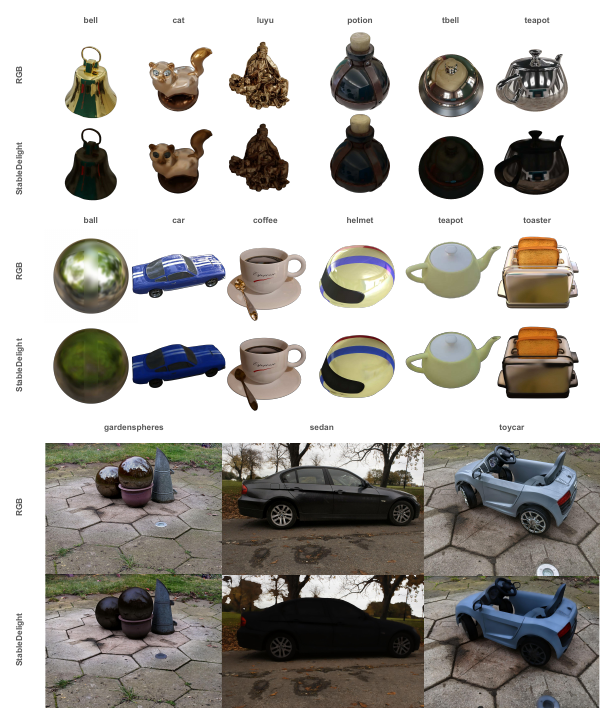}
    \caption{Example predictions of StableDelight~\cite{stabledelight2025} on scenes from the Shiny Synthetic \cite{verbin2022refnerf}, Glossy Synthetic \cite{liu2023nero}, and Shiny Real \cite{verbin2022refnerf} datasets.}
    \label{fig:supplementary:stabledelight}
\end{figure*}

\begin{figure*}
    \centering
    \includegraphics[width=\linewidth]{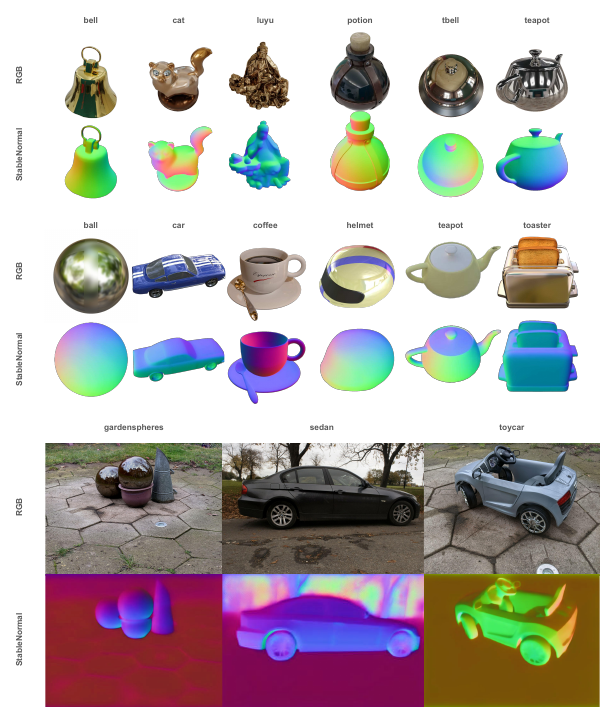}
    \caption{Example predictions of StableNormal~\cite{ye2024stablenormal} on scenes from the Shiny Synthetic \cite{verbin2022refnerf}, Glossy Synthetic \cite{liu2023nero}, and Shiny Real \cite{verbin2022refnerf} datasets.}
    \label{fig:supplementary:stablenormal}
\end{figure*}

\cref{fig:supplementary:stabledelight} and \cref{fig:supplementary:stablenormal} present sample predictions of StableDelight~\cite{stabledelight2025} and StableNormal~\cite{ye2024stablenormal} for all examined scenes from the Shiny Synthetic~\cite{verbin2022refnerf}, Glossy Synthetic~\cite{liu2023nero} and Shiny Real \cite{verbin2022refnerf} scenes. StableDelight manages to remove specular reflections in most cases, but fails in cases of mirror-like surfaces (e.g. toaster and glossy teapot), confusing the reflections as part of the diffuse appearance of the scene. StableNormal, on the other hand, fails to recognize the surface of the liquid in the coffee scene and oversmooths complex geometries (e.g. cat, luyu) and occasionally confuses interreflections as separate geometries (e.g. tbell).

\end{document}